\documentclass[pdflatex,twocolumn,english,prb,aps,showpacs,superscriptaddress]{revtex4-1}
\usepackage{multirow}
\usepackage{tikz}%tikzの読み込み
\usepackage{pxpgfmark}%tikzのnode名を共有

\usetikzlibrary{arrows,positioning,plotmarks,external,patterns,angles,
decorations.pathmorphing,backgrounds,fit,shapes}
\usetikzlibrary{shapes.callouts}

\usepackage{amsmath}%数式
\usepackage{amssymb}%数式
\usepackage{bm}%数式太字
\usepackage{txfonts}
\usepackage{newtxmath}%数式文字
\usepackage{newtxtext}%テキスト文字
\usepackage{graphicx}%図
\usepackage{physics}%物理系のマクロ
\usepackage{float}
\usepackage{hyperref}
\usepackage{algorithm}
\usepackage{algpseudocode}
\usepackage{mathtools}%数学の定義の記号を使うためのパッケージ
\usepackage{braket}

\usepackage{pgfplots}
\pgfplotsset{compat=1.12}
\makeatletter
\renewcommand*\env@matrix[1][*\c@MaxMatrixCols c]{%
  \hskip -\arraycolsep
  \let\@ifnextchar\new@ifnextchar
  \array{#1}}
  \makeatother%}}}

\usepackage{booktabs}
\newcommand{\mebius}{\hspace{0pt}_{\scalebox{1.5}{$\bullet$}} }
\newcommand{\ahat}{\hat{a}}
\newcommand{\bhat}{\hat{b}}

\newcommand{\mathcalz}{\mathcal{Z}}

\newcommand{\rhoj}{\rho\left(E, j\right)}

\newcommand{\rmax}{r_{\mathrm{max}} }

\newcommand\eq[1]{Eq.~(#1)}
\newcommand\eqs[1]{Eqs.~(#1)}
\newcommand\headeq[1]{Equation~(#1)}
\newcommand\headeqs[1]{Equations~(#1)}
\newcommand\tab[1]{Table~#1}
\newcommand\fig[1]{Fig.~#1}
\newcommand\figs[1]{Figs~#1}
\newcommand\headfig[1]{Figure~#1}
\newcommand\headfigs[1]{Figures~#1}
\newcommand\sect[1]{Sect.~#1}
\newcommand\sects[1]{Sects.~#1}

\newcommand\headapp[1]{Appendix~#1}
\newcommand\headalg[1]{Algorithm~#1}

\RequirePackage{xcolor}
\RequirePackage{tikz}
\definecolor{mygray}{HTML}{717171}
\definecolor{myred}{HTML}{C7243A}
\definecolor{myblue}{HTML}{3261AB}
\definecolor{mygreen}{HTML}{009250}
\definecolor{myorange}{HTML}{ED8917}
\definecolor{mypurple}{HTML}{C23685}
\definecolor{myviolet}{HTML}{988ACA}
\usepackage{ulem}
\usepackage{cancel}
\begin{document}

% \title{Theory of the Spatial Dependence of Odd-Frequency Pairing in 
% Kitaev Chain beyond Quasiclassical Approximation}
\title{Odd-frequency pairing and proximity effect in Kitaev chain systems\\ including topological critical point}

\author{Daijiro Takagi}
\affiliation{Department of Applied Physics, Nagoya University, Nagoya 464-8603, Japan}

\author{Shun Tamura}
\affiliation{Department of Applied Physics, Nagoya University, Nagoya 464-8603, Japan}

\author{Yukio Tanaka}
\affiliation{Department of Applied Physics, Nagoya University, Nagoya 464-8603, Japan}

\begin{abstract}
    In this paper, we investigate the relation between odd-frequency pairing and proximity effect in non-uniform Kitaev chain systems with a particular interest in the topological critical point.  First, we correlate the odd-frequency pairing and Majorana fermion in a semi-infinite Kitaev chain, where we find that the spatial dependence of the odd-frequency pair amplitude coincides with that of the local density of states at low frequencies. Second, we demonstrate that, contrary to the standard view, the odd-frequency pair amplitude spreads into the bulk of a semi-infinite Kitaev chain at the topological critical point. Finally, we show that odd-frequency Cooper pairs cause the proximity effect in a normal metal/diffusive normal metal/ Kitaev chain junction even at the topological critical point. Our results hold relevance to the investigation of odd-frequency pairing and topological superconductivity in more complicated systems that involve Rashba nanowire with magnetic fields.
\end{abstract}

%%% Keywords are not needed any longer. %%%
%%%\kword{keyword1, keyword2, keyword3, \ldots}
%%%

\maketitle

\section{Introduction}
Superconductors are characterized by Cooper pairs which result from the pairing between electrons. 
They are described by pair amplitude that is antisymmetric due to Fermi-Dirac statistics. 
These statistics allow the pair amplitude to be an even function of the time coordinates, or even-frequency in the frequency domain, of the paired electrons. 
In its most general form, however, Fermi-Dirac statistics also allows the pair amplitude to be odd under the exchange of time coordinates, or odd-frequency in the frequency domain\cite{Berezinskii}.

The realization of odd-frequency pairing has been predicted to occur in bulk systems, although experimental demonstration has remained so far elusive\cite{Belitz1,Balatsky,Kusunose,Fominov2015}. 
On the other hand, odd-frequency pairing can be induced from even-frequency superconductors due to translational symmetry breaking\cite{odd3,tanaka12} (in non-uniform systems) and spin-rotational symmetry breaking\cite{Efetov1,Efetov2,buzdin_rmp,eschrig2015spin,Golubov_RMP,linder2015superconducting} (in systems with exchange fields). 
Remarkable phenomena caused by odd-frequency pairs include the long-range proximity effect in an $s$-wave superconductor/ferromagnet junction\cite{Efetov1,Efetov2,buzdin_rmp,eschrig2015spin,Golubov_RMP,linder2015superconducting}, the paramagnetic response\cite{Meissner10,SuzukiAsano1,SuzukiAsano2,Lutchyn2017,Meissner3,Bernardo2,Asano2011}, and the anomalous proximity effect\cite{odd1,Proximityp,Proximityp2,Proximityp3,Ikegaya2016}.
Recently, two of the authors have shown that odd-frequency pairs, appearing in topological systems, have a singularity of $1/\omega$, where $ \omega $ is frequency\cite{TamuraHoshino}.
In this paper, we focus on the anomalous proximity effect caused by this special odd-frequency pairs.

In junctions with unconventional superconductors such as spin-singlet $d$-wave or spin-triplet $p$-wave\cite{ABS,ABSb,Hu94,TK95,ABSR1,odd3,tanaka12}, the odd-frequency pairing can be enhanced due to the presence of the zero-energy surface Andreev bound states. 
However, the odd-frequency spin-triplet $s$-wave pairs, appearing in junctions with spin-triplet superconductors such as $p$-wave or $f$-wave, can penetrate the diffusive normal metal region\cite{odd1,Proximityp,Proximityp2,Proximityp3,Ikegaya2016}. 
This is the so-called anomalous proximity effect.
As results of this effect, a zero-energy peak of the local density of states occurs in the diffusive normal metal region, and a zero-bias conductance is quantized\cite{Proximityp,Proximityp2,Proximityp3,Ikegaya2016}.
This effect is in contrast to the conventional proximity effect realized in junctions with spin-singlet $s$-wave superconductors\cite{Golubov88}.

Experimentally, the anomalous proximity effect has not been observed yet.
The most straightforward system for investigating the effect is a diffusive normal metal/$p$-wave superconductor junction\cite{Proximityp,Proximityp2,Proximityp3}.
Actually, no $p$-wave superconductor has been found yet.

Hence, many attempts have been made to create a system like a $p$-wave superconductor from an $s$-wave superconductor\cite{STF09,STF10,lutchyn10,oreg10}.
A typical example is a semiconductor nanowire, with Rashba spin-orbit coupling and magnetic fields, stacked on the $s$-wave superconductor\cite{lutchyn10,oreg10}.
Moreover, it is shown that the low-energy excitation in the system is equivalent to that in the $p$-wave superconductor\cite{Asano2013}.
In several experiments, the zero energy peaks of the local density of states have been observed\cite{Mourik,zhang2018quantized}.

Theoretically, the proximity effect of unconventional superconductors has been studied based on the so-called quasiclassical approximation \cite{Serene,Eilenberger,Eschrig00}. 
In this approach, we integrate the energy scale apart from the Fermi energy $E_{F}$ and  the length scale of the physical quantity is governed by the coherence length $\xi=\hbar k_{F}/|\Delta|$ with pair potential $\Delta$ in a superconductor.
This approximation is justified for $k_{F} \xi \geq 1$. 
The quasiclassical approximation has been applied for non-uniform unconventional superconductors and superfluid $^{3}$He. 
With the quasiclassical approximation, many essential concepts like the zero-energy surface Andreev bound states are successfully derived \cite{ABS,ABSb,Hu94,TK95,ABSR1}. 
In this regard, the theory employed to investigate the anomalous proximity effect has been based on this quasiclassical approximation\cite{Proximityp,Proximityp2,Proximityp3}.

However, quite often, it is indispensable to characterize the properties of $p$-wave junctions beyond the quasiclassical approximation.
This is because the regime, where quasiclassical approximation does not work, is achieved by controlling external parameters.
The control of parameters can change the topological non-trivial regime (the topological regime) with the zero-energy surface Andreev bound states\cite{TK95,ABSR1} to the trivial regime (the non-topological regime) without the zero-energy Andreev bound states in the $p$-wave superconductor junction.

The minimal model to express a spin-triplet $p$-wave superconductor covering from the topological to the non-topological regime is the Kitaev model which consists of a one-dimensional chain of spinless fermions \cite{Kitaev01}. 
In this model, the chemical potential drives a transition between the topological (metallic) to the non-topological (insulating) regimes.
The quasiclassical theory is available only for the topological regime far from the topological critical point which is the border between the topological and non-topological phases. 
It is a challenging problem to study the odd-frequency pairing realized in this model beyond quasiclassical approximation for the reasons explained in the previous paragraph.
In this paper, we investigate the odd-frequency pairing beyond the quasiclassical approximation in junctions based on the Kitaev chain by applying the recursive Green's function method \cite{umerski1997closed}.

First, we address the spatial dependence of the odd-frequency pairing in the semi-infinite Kitaev chain.
In the topological regime far from the topological critical point, $k_{F} \xi \geq 1$ is satisfied. 
However, near the topological critical point, this relation does not hold anymore and the scale of the coherence length becomes the same order of $1/k_{F}$. 
We demonstrate the change of the spatial dependence of the odd-frequency pairing towards the topological critical point. 
Since the Majorana fermion is strictly relevant to the odd-frequency pairing \cite{Asano2013},  we compare the spatial dependence of the odd-frequency spin-triplet pair amplitude at low frequency, that of the local density of states at zero-energy, and that of the wave function of Majorana fermion. 
We find that the odd-frequency pair amplitude, localized at the edge in the topological regime, increases and spreads into the bulk regime and takes a constant value in bulk at the topological critical point.

Second, we study the anomalous proximity effect in a normal metal/diffusive normal metal/Kitaev chain junction. 
In particular, we focus on the spatial dependence of the odd-frequency spin-triplet $s$-wave pair amplitude both in the diffusive normal metal region and Kitaev chain region near the topological critical point and compare it to the even-frequency spin-triplet $p$-wave pair amplitude. We demonstrate that the magnitude of the odd-frequency pair amplitude has a maximum in the Kitaev chain region near the diffusive normal metal/Kitaev chain boundary and it decreases towards the topological critical point.

We also calculate the local density of states of quasiparticle in the diffusive normal metal region and zero bias voltage conductance. 
In the topological regime, the local density of states has a zero-energy peak and a quantized zero-bias conductance peak. We study the peak width of the zero-bias conductance peak as a function of the length of $L$ and the strength of impurity potential in the diffusive normal metal region. 
Just at the topological critical point, the local density of states still has a zero-energy peak; however, the zero-bias conductance peak disappears and the  conductance at zero bias voltage is no more quantized.
Since the topological critical point corresponds to the metal-insulator transition point, to show the nature of the proximity effect around the metal-insulator transition point, we also calculate the spatial dependence of the even-frequency spin-singlet $s$-wave pair amplitude in a normal metal/diffusive normal metal/$s$-wave superconductor junction around the metal-insulator transition point where the conventional proximity effect occurs. 

This paper is organized as follows.
In \sects{\ref{sect:kitaev}--\ref{sect:method}}, we introduce the Kitaev chain and method.
In \sect{\ref{sect:semi-infinite}}, we calculate the spatial dependence of odd-frequency pairing in the semi-infinite Kitaev chain.
In \sect{\ref{sect:junction}}, we study the proximity effect in the regime including the topological critical point in the normal metal/diffusive normal metal/Kitaev chain junction.

\section{Kitaev Chain}
\label{sect:kitaev}
We consider inhomogeneous systems based on the Kitaev chain, which models 1D $p$-wave superconductivity. It can be described by the following Hamiltonian,
\begin{align}
    \label{eq:kitaev}
    \mathcal{H} &= - t \sum_{j} \left( c_j^\dagger c_{j+1} + c_{j+1}^\dagger c_j \right) - \mu \sum_{j} c_j^\dagger c_j \nonumber\\
                &\quad + \sum_{j} \left( \Delta c_j^\dagger c_{j+1}^\dagger + \mathrm{H.c.} \right),
\end{align}
where $c_{j}^\dagger$($c_{j}$), $t$, $\mu$, and $\Delta$ are 
a creation (an annihilation) operator at site $j$, hopping between nearest neighboring sites, chemical potential measured from Fermi energy, and pair potential, respectively,
and $\mathrm{H.c.}$ denotes Hermitian conjugate.
The macroscopic phase of the pair potential is chosen to be zero.
\headeq{\ref{eq:kitaev}} does not include the spin index because we can regard the Kitaev chain as a spinless state by considering a fully polarized spin-triplet pairing like $\ket{\uparrow\uparrow}$ 
($\ket{\downarrow\downarrow}$) spin state.

\begin{figure}[!tbp]
    \centering
    \includegraphics[scale = 0.7]{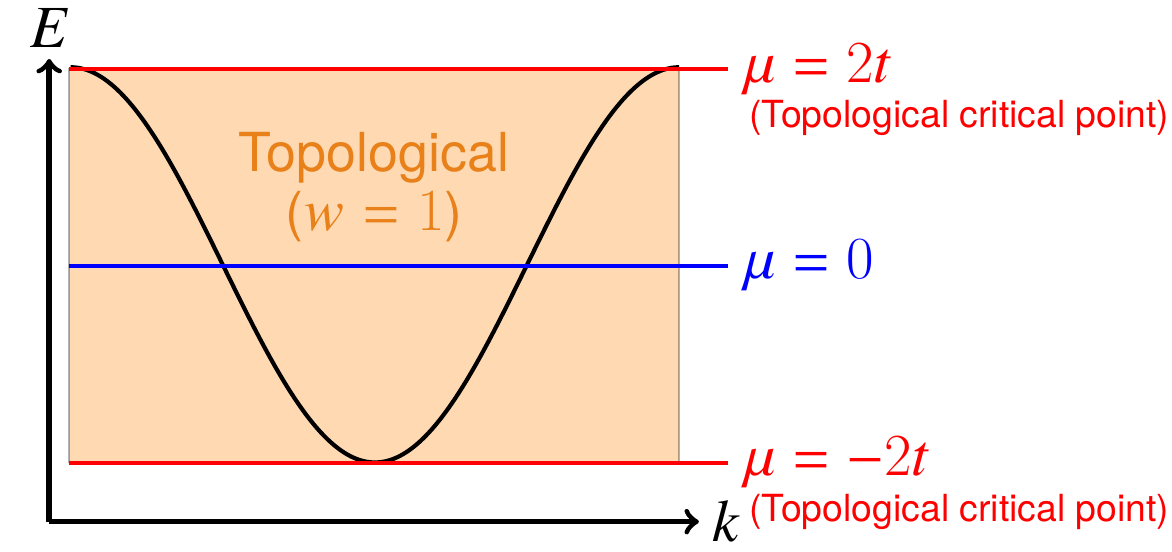}
    \caption{
        The energy dispersion of the Kitaev chain in the normal state with  $\Delta=0$.
        When a chemical potential is in the energy band $-2t < \mu < 2t$, 
        a topological phase with Majorana fermion is realized.
        At the topological critical point, $\mu$ is located at the edge of the energy band.
    }
    \label{fg:band}
\end{figure}

In the Kitaev chain, a topological phase can be expressed 
only by using the information of the energy band in a normal state {[\fig{\ref{fg:band}}]}.
In the topological regime $|\mu|<2t$,
the chemical potential is located on the energy band.
In the non-topological regime $|\mu|>2t$,
the Kitaev chain in normal states becomes an insulator.
Here, we identify topological critical points at $\mu=\pm2t$, which are located at the boundaries between the topological and non-topological regime.

It has been demonstrated that  the Kitaev model has a topological invariant, winding number w, given by,\cite{Sumanta,TamuraHoshino}
\begin{align}
    w &=\frac{-1}{4\pi i}\int_{-\pi}^{\pi}
    \mathrm{Tr}\left[\Gamma{H}^{-1}(k)\partial_kH(k)\right] 
    \mathrm{d}k
    =
    \left\{
    \begin{array}{cl}
        1 & (|\mu|<2t)\\
        0 & (|\mu|>2t)
    \end{array}\right.,
\end{align}
where $\Gamma=\tau_x$ is a chiral operator satisfying
$\{\Gamma,H(k)\}=0$ and $\tau_x$ is a Pauli matrix.
$H(k)$ is defined as $\mathcal{H}=(1/2)\sum_k[c_k^\dagger,c_{-k}] H(k) [c_k,c_{-k}^\dagger]^\mathrm{T}$ where $k$ and  $c_k$ are a wave number and the Fourier transformed form of $c_j$ into momentum space, respectively.
The winding number $w$ is zero for the non-topological phase and $1$ for the topological phase.
We stress that the winding number cannot be defined at the topological critical points since the system at such points is gapless.
Hence, the value of $w$ exhibits a discontinuous jump.

\section{Method}
\label{sect:method}
In this section, we introduce the Green's function method which is used to calculate the pair amplitudes and various physical quantities numerically. 
In particular we employ the retarded (advanced) Green's functions in real space to be calculated as \cite{Ebisu}
\begin{align}
    \label{eq:green}
    \check{G}^{R(A)}(E,j,j') &= \left\{\left[\left(E +(-) i\delta_\epsilon\right)I - H\right]^{-1}\right\}_{j,j'}\nonumber\\
    &=
    \begin{bmatrix}
        G_{j, j'}^{R(A)}(E)         & F_{j, j'}^{R(A)}(E) \\
        \tilde{F}_{j, j'}^{R(A)}(E) & \tilde{G}_{j, j'}^{R(A)}(E)
    \end{bmatrix},
\end{align}
where $\check{\hspace{0.5em}}$ denotes a $2\times2$ matrix, $E$ is 
quasiparticle energy measured from the chemical potential,
$\delta_\epsilon$ is an infinitesimal positive number, $H$ is defined as $\mathcal{H} = (1/2)[\ldots, c_1^\dagger, c_1, c_2^\dagger, c_2, \ldots]H[\ldots, c_1, c_1^\dagger, c_2, c_2^\dagger, \ldots]^\mathrm{T}$, 
and $I$ is an identity matrix with the same size as $H$.
    In the second equality of \eq{\ref{eq:green}}, we made use of electron-hole symmetry, where $G$ and $F$ correspond to the normal electron-electron and anomalous electron-hole components of the Green's function $\check{G}$ in \eq{\ref{eq:green}}.
    In this work, we make use of \eq{\ref{eq:green}} to investigate the induced odd- and even-frequency pair amplitudes, the local density of states, and differential conductance. 

    The local density of states is obtained from the following expression:
\begin{align}
    \label{eq:ldos}
    \rho(E,j) = - \frac{1}{\pi}\mathrm{Im}\left[G_{j,j}^R(E)\right],
\end{align}
with $G_{j,j}^{R}$ given by \eq{\ref{eq:green}}. 

In a normal metal/diffusive normal metal/Kitaev chain, the Lee-Fisher formula \cite{lee1981anderson} allows us to calculate the differential conductance:
\begin{align}
    \label{eq:cond}
    G_\mathrm{NS}(E, j) = \frac{t^2e^2}{2h} \mathrm{Tr} & \left[-\bar{G}_{j,j+1}\bar{G}_{j,j+1} - \bar{G}_{j+1,j}\bar{G}_{j+1,j} \right.\nonumber\\
                                                               &\quad \left. + \bar{G}_{j,j} \bar{G}_{j+1,j+1} + \bar{G}_{j+1,j+1}\bar{G}_{j,j} \right],
\end{align}
where $\bar{G}_{j,j'} = (G_{j,j'}^A - G_{j,j'}^R)/2i$. 

The pair amplitudes are investigated from Matsubara Green's functions which can be obtained from \eq{\ref{eq:green}} by replacing $E+(-)i\delta\to i\omega_n$, where $\omega_n$ is Matsubara frequency. According to Fermi-Dirac statistics, the pair amplitudes obey,  
\begin{align}
    \label{eq:odd cooper pairs}
    \tilde{F}_{jj^\prime\sigma\sigma^\prime}(i\omega_n)=-\tilde{F}_{j^\prime j\sigma^\prime\sigma}(-i\omega_n),
\end{align}
where $\sigma$ and $\sigma^\prime$ are spin of two electrons forming a Cooper pair.
\headeq{\ref{eq:odd cooper pairs}} 
allows four types of symmetries for the pair amplitudes: even-frequency spin-singlet even-parity (ESE), odd-frequency spin-triplet even-parity (OTE)\cite{Berezinskii}, even-frequency spin-triplet odd-parity (ETO), and odd-frequency spin-singlet odd-parity (OSO)\cite{Balatsky}.
Kitaev chain has odd-frequency spin-triplet even-parity and even-frequency spin-triplet odd-parity pair amplitude because it
represents a chain of spin-polarized electrons, which corresponds to spin-triplet, with $p$-wave pairing.
In the Kitaev chain, the odd- and even-frequency pair amplitudes are given as,
\begin{align}
    \label{eq:ote}
    f^{\mathrm{OTE}}(j)
    &= \frac{1}{2}\left[\tilde{F}_{j,j}(i\omega_n)-\tilde{F}_{j,j}(-i\omega_n)\right],\\
    \label{eq:eto}
    f^\mathrm{ETO} (j)
    &= \frac{1}{2}\left[\frac{\tilde{F}_{j+1, j}(i\omega_n)-\tilde{F}_{j, j+1}(i\omega_n)}{2}\right.\nonumber\\
    &\qquad+\left.\frac{\tilde{F}_{j+1, j}(-i\omega_n)-\tilde{F}_{j, j+1}(-i\omega_n)}{2}\right],
\end{align}
respectively.
As a comparison, we also calculate the spin-singlet $s$-wave superconductor 
junction. 
Unlike the Kitaev chain, an $s$-wave superconductor has a spin degree of freedom.
Therefore, the anomalous Green's function $\tilde{F}_{j,j'}(i\omega_n)$ becomes $\tilde{F}_{j,j',\uparrow,\downarrow}(i\omega_n)$.
The even-frequency spin-singlet even-parity pair amplitude is given as below\cite{Ebisu}:
\begin{align}
    \label{eq:ese}
    f_{\uparrow\downarrow-\downarrow\uparrow}^{\mathrm{ESE}}(j)
    &= \frac{1}{2}\left[\tilde{F}_{j,j,\uparrow,\downarrow}(i\omega_n)+\tilde{F}_{j,j,\uparrow,\downarrow}(-i\omega_n)\right].
\end{align}

\section{Semi-infinite Kitaev Chain}
\label{sect:semi-infinite}
\begin{figure}[!tbp]
    \centering
    \includegraphics[scale = 0.8]{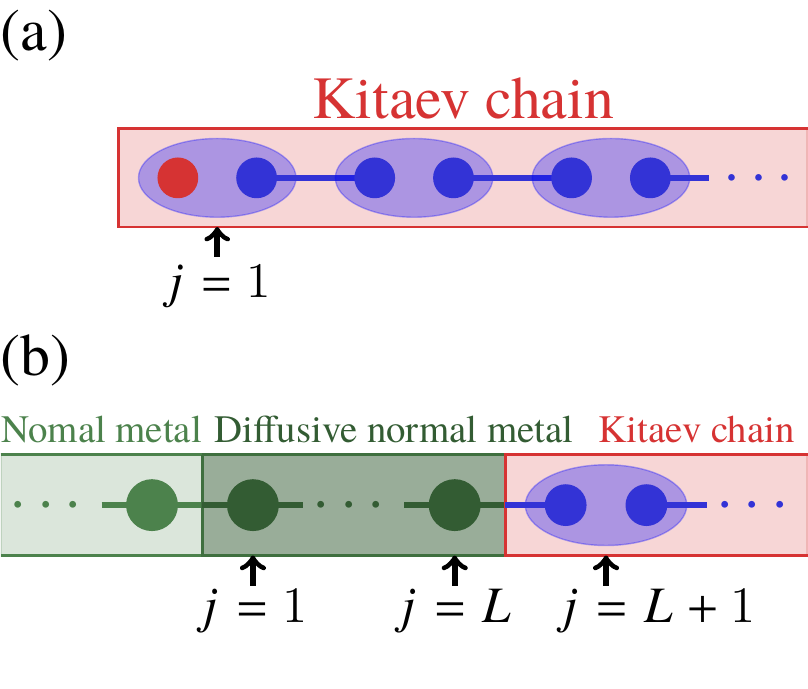}
    \caption{
        (a) Semi-infinite system of Kitaev chain;
        (b) Normal metal/diffusive normal metal/Kitaev chain junction. $j$: the  number of sites in the chain,
        $j\le0$: a ballistic metal without impurities, 
        $1\le j \le L$: diffusive normal metal containing impurities,
        $L+1\le j$: Kitaev chain,
        $L$: the number of sites in the diffusive normal metal region.
    }
    \label{fg:system}
\end{figure}
In this section, we consider a semi-infinite Kitaev chain, as depicted in \fig{\ref{fg:system}}(a).
We investigate the spatial dependence of odd-frequency correlations and its relation to Majorana fermion.
It has been pointed out that Majorana fermion accompanies odd-frequency pairing\cite{Asano2013}. 
However, their spatial dependence has not been addressed yet.
The relation is derived within the recursive Green's function approach, and the details of the calculation can be found in \headapp{\ref{sub:recursive}}.
\begin{figure}[!tbp]
    \centering
    \includegraphics[scale = 0.5]{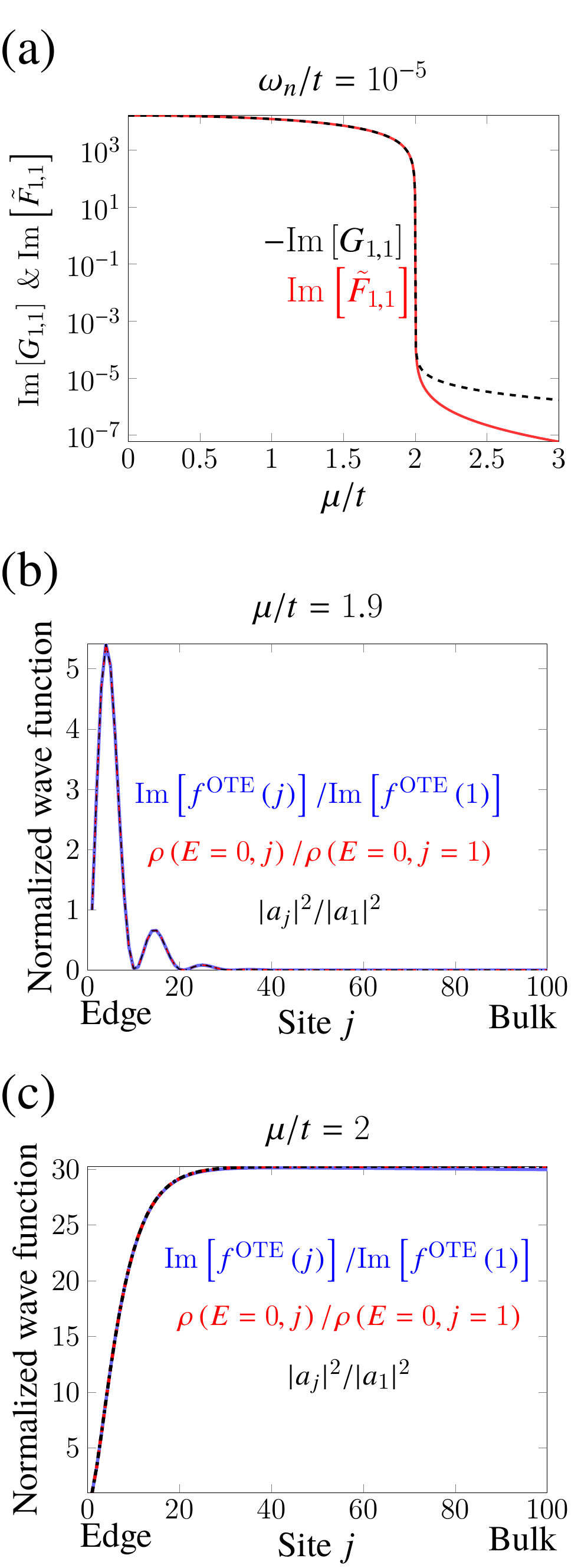}
    \caption{
        (a) Imaginary part of normal Green's function $\mathrm{Im}[G_{1,1}]$
        and anomalous Green's function $\mathrm{Im}[\tilde{F}_{1,1}]$
        are plotted as a function of $\mu$ in the semi-infinite Kitaev chain.
        $\omega_n$: Matsubara frequency,
        $\mathrm{Im}[\tilde{F}_{1,1}]$ 
        : the odd-frequency spin-triplet even-parity pair amplitude. 
        In the low $\omega_{n}$ limit,  
        $-\mathrm{Im}[G_{1,1}]$: the local density of states of Majorana fermion in topological phase with $|\mu|<2t$.
        $\omega_n/t=10^{-5}$;
        (b,c) The site dependence of probability density of Majorana wave function $|a_j|^2$, 
        odd-frequency spin-triplet even-parity pair amplitude $\mathrm{Im}[f^\mathrm{OTE}(j)]$ with $\omega_n/t=10^{-5}$, 
        and the local density of states $\rho(E=0,j)$ with $\delta_\epsilon/t=10^{-5}$.
        All of these quantities are calculated in the semi-infinite Kitaev chain and are normalized by their values at $j=1$.
        (b) $\mu/t=1.9$; (c) $\mu/t=2$;
        $\Delta/t=0.1$.
    }
    \label{fg:semi-inf}
\end{figure}

\headfig{\ref{fg:semi-inf}(a)} shows the $\mu$ dependence between the normal and anomalous components of Green's functions [\eq{\ref{eq:green}}] at the edge of the system ($j=1$) using numerical calculation.
The imaginary part of normal Green's functions, $\mathrm{Im}[G_{1,1}]$, corresponds to the local density of states of Majorana fermion;
that of anomalous Green's functions, $\mathrm{Im}[F_{1,1}]$, stands for the imaginary of the odd-frequency pair amplitude. 
Our result indicates that $-\mathrm{Im}[G_{1,1}] = \mathrm{Im}[F_{1,1}]$ in the topological regime for $\omega_n/t = 10^{-5}$ as shown in \fig{\ref{fg:semi-inf}(a)}.
In particular, the relative error between $\mathrm{Im}[G_{1,1}]$ and $\mathrm{Im}[\tilde{F}_{1,1}]$
is $10^{-9}$ for $\mu=0$.
As observed in \fig{\ref{fg:semi-inf}(a)}, $-\mathrm{Im}[G_{1,1}]$ and $\mathrm{Im}[F_{1,1}]$ take a maximum for $\mu=0$ and decrease as $\mu$ increases. 
As $\mu$ approaches the topological critical point, which correspond to $\mu=2$ in \fig{\ref{fg:semi-inf}}, these value are suppressed dramatically. 
In the non-topological regime, these values are small ($-\mathrm{Im}[G_{1,1}]\sim10^{-5}$ and $\mathrm{Im}[F_{1,1}]\sim10^{-7}$, respectively).

The discussion above can be further analytically supported by evaluating the following expression,
\begin{align}
    \label{eq:majo-odd-ana}
    &\lim_{\omega_n\to0}\left\{\mathrm{Im}\left[G_{1,1}\right] + \mathrm{Im}\left[\tilde{F}_{1,1}\right]\right\}\nonumber\\
    &\quad\simeq \lim_{\omega_n\to0}\left\{
        \begin{array}{ll}
            \frac{-\omega_n}{\omega_n^2+4t^2} & \mbox{(Topological)}\\
            \frac{-\omega_n}{\omega_n^2+\mu^2} & \mbox{(Non-topological)}
        \end{array}
        \right.
        =0
        ,
\end{align}
which has been evaluated at $\Delta=t$ in the zero-frequency limit $\omega_n\to0$. 
(See \headapp{\ref{sec:proof-delta-t}} for the detailed deviation of \eq{\ref{eq:majo-odd-ana}}).
    From \eq{\ref{eq:majo-odd-ana}} we conclude that in the topological regime, $\mathrm{Im}[G_{1,1}] + \mathrm{Im}[\tilde{F}_{1,1}]$ depends on $t$ while it depends on $\mu$ in the non-topological regime at low-frequency. In the non-topological regime, $\mathrm{Im}[G_{1,1}] + \mathrm{Im}[\tilde{F}_{1,1}]$ converges faster to zero as $\mu$ increases.

    Furthermore, we are interested in the Majorana wave function, which has been derived before\cite{hegde2016majorana} and reads,
\begin{align}
    \label{eq:majowave}
    a_j &= a_1 C^{j-1} \left\{\cos{\left[\beta (j-1)\right]}+ \frac{1}{\tan{\beta}} \sin{\left[\beta (j-1)\right]}\right\},
\end{align}
where $C=\sqrt{t-\Delta}/\sqrt{t+\Delta}$ and $\beta = \arctan{(\sqrt{4t^2-4\Delta^2-\mu^2}/\mu)}$.
In \figs{\ref{fg:semi-inf}(b,c)}, we
compare the spatial dependence
between the probability density $|a_j|^2$ of Majorana wave function,
the local density of states $\rho(E=0,j)$ {[\eq{\ref{eq:ldos}}]}, 
and the odd-frequency pair amplitude $\mathrm{Im}[f^\mathrm{OTE}(j)]$ {[\eq{\ref{eq:ote}}]}.
These three values are normalized by their values  at $j=1$.

In \fig{\ref{fg:semi-inf}(b)}, these three values coincide with each other in the topological regime near the topological critical point.
The spatial oscillations of them originate from Friedel oscillation 
with the length scale $1/k_{F}$
because the existence of the boundary is a big perturbation to bulk condensation.
The oscillation appears for $(\mu/2)^2+\Delta^2<t^2$ that is derived by Hegde \textit{et al.}\cite{hegde2016majorana}.
Notice that, even at the topological critical point, these three values coincide with each other [\fig{\ref{fg:semi-inf}(c)}] but their behavior exhibits a dramatic different dependence from that in the topological regime [\fig{\ref{fg:semi-inf}(b)}].
There is no oscillatory behavior and they increase as a function of $j$ and becomes constant for $j>40$.

It has been thought that the odd-frequency spin-triplet even-parity pairing is localized at the edge.
However, it is a surprising effect that the odd-frequency spin-triplet even-parity pairing 
exists in the Kitaev chain region far from the edge as shown in \fig{\ref{fg:semi-inf}(c)}.
This result suggests the possibility of realizing odd-frequency superconductivity in bulk at the topological critical point in the semi-infinite system.

Now we explain for why the odd-frequency spin-triplet even-parity pairing still exists in bulk.
For $|\mu|\leq2t$, $0<\Delta<t$, and $\omega_n/t = 10^{-5}$,
the spatial dependence of the odd-frequency pair amplitude, based on \eq{\ref{eq:majowave}}, is obtained as
\begin{align}
    \label{eq:odd-analytical}
    \mathrm{Im}\left[f^\mathrm{OTE}(j)\right]  &= \mathrm{Im}\left[f^\mathrm{OTE}(1)\right] C^{2j-2} \nonumber\\
            &\quad\times\left|\cos{\left[\beta (j-1)\right]}+ \frac{1}{\tan{\beta}} \sin{\left[\beta (j-1)\right]}\right|^2.
\end{align}

To understand the exotic spatial dependence of the odd-frequency spin-triplet even-parity pair amplitude at the topological critical point, 
we choose  $\mu/t=2$ and assume $\Delta/t=\alpha$ $(0<\alpha<1)$ and $j\gg1$. 
Then, we can express \eq{\ref{eq:odd-analytical}} as below:
\begin{align}
    \label{eq:odd-analytical-j-inf}
    \frac{\mathrm{Im}\left[f^\mathrm{OTE}(j)\right]}{\mathrm{Im}\left[f^\mathrm{OTE}(1)\right]} &\simeq \left(\frac{1+\alpha}{2\alpha}\right)^2,
\end{align}
where $\beta = \arctan(i\alpha)$.
\headeq{\ref{eq:odd-analytical}} shows that the $\mathrm{Im}[f^\mathrm{OTE}(j)]$ is independent of $j$ for $\mu/t=2$ and $j\gg1$.
It is noted that we can analytically show that the odd-frequency spin-triplet even-parity pairing takes a constant value far from the edge at the topological critical point.
By substituting $\alpha=0.1$ for \eq{\ref{eq:odd-analytical-j-inf}}, 
$\mathrm{Im}[f^\mathrm{OTE}(j)]/\mathrm{Im}[f^\mathrm{OTE}(1)]\to30.25$ is confirmed 
as shown in \fig{\ref{fg:semi-inf}(c)}.
This result is remarkable and overturns the previous view that odd-frequency pairs are localized at the end of the system.

\section{Normal Metal/Diffusive Normal Metal/Kitaev Chain Junction}
\label{sect:junction}

\begin{figure*}[!htbp]
    \centering
    \includegraphics[scale=0.4]{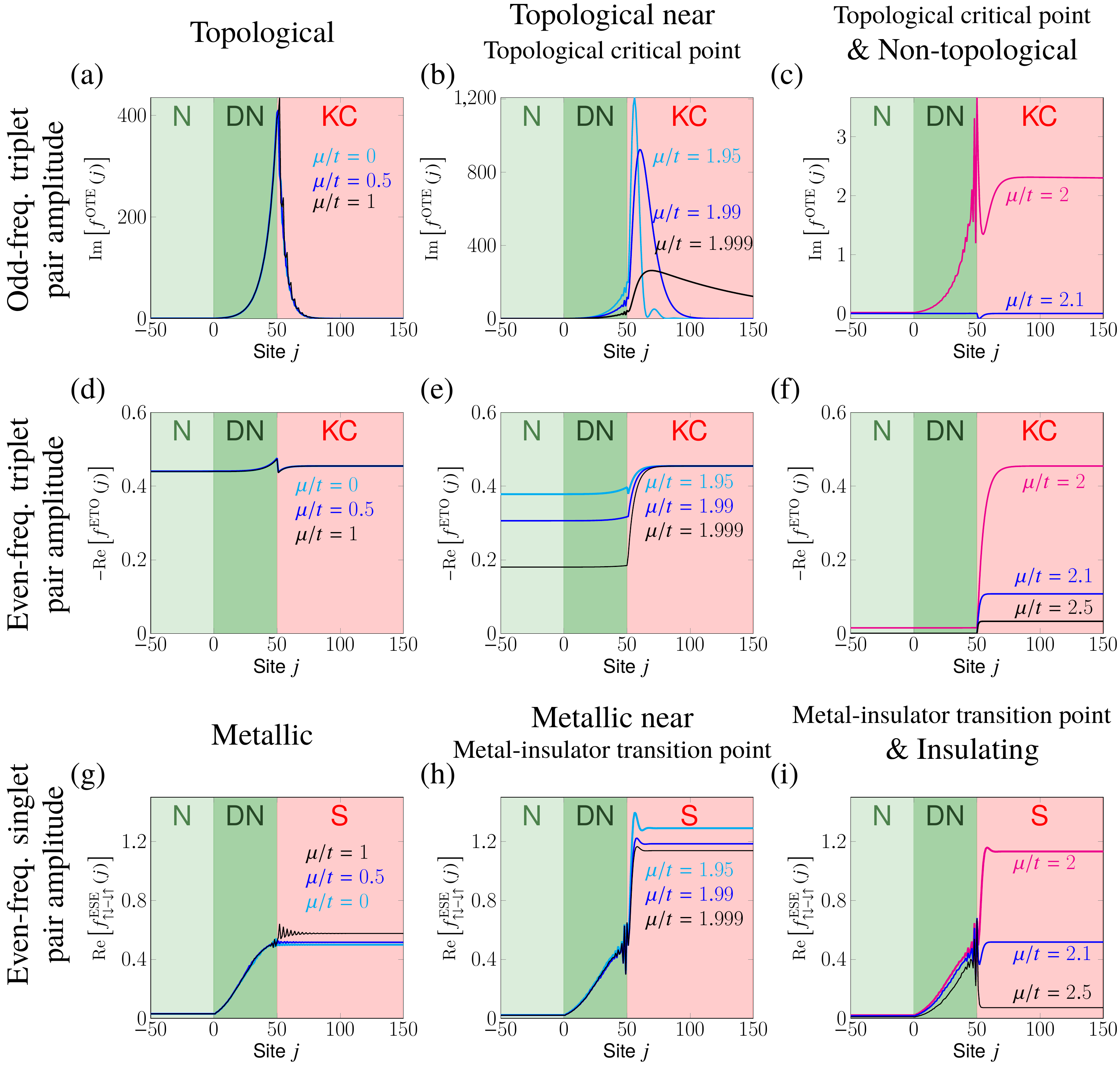}
    \caption{
        (a--c) odd-frequency spin-triplet even-parity (OTE) pair amplitude in the normal metal/diffusive normal metal/Kitaev chain (N/DN/KC) junction;
        (d--f) even-frequency spin-triplet odd-parity (ETO) pair amplitude in the normal metal/diffusive normal metal/Kitaev chain junction;
        (g--i) even-frequency spin-singlet even-parity (ESE) pair amplitude in the normal metal/diffusive normal metal/$s$-wave superconductor (N/DN/S) junction;
        $\mu_\mathrm{N}/t=0.5$, $\Delta/t=0.1$, $\omega_n/t=10^{-5}$, $L=50$, and $W/t=3$.
        (a,d) topological regime: $\mu/t=0$, $0.5$, $1$;
        (b,e) topological regime near topological critical point: $\mu/t=1.95$, $1.99$, $1.999$;
        (c,f) topological critical point: $\mu/t=2$; non-topological regime: $\mu=2.1$, $2.5$.
        (g) metallic regime: $\mu/t=0$, $0.5$, $1$;
        (h) metallic regime near metal-insulator transition point: $\mu/t=1.95$, $1.99$, $1.999$;
        (i) metal-insulator transition point: $\mu/t=2$; insulating regime: $\mu=2.1$, $2.5$.    
    }
    \label{fg:pair-amp}
\end{figure*}

In this section, we consider a normal metal/diffusive normal metal/Kitaev chain junction as depicted in \fig{\ref{fg:system}(b)} and study the proximity effect.
The Hamiltonian of the system is given by
\begin{align}
    \label{eq:hamiltonian-junction}
    \mathcal{H} &= - t \sum_j \left( c_j^\dagger c_{j+1} + c_{j+1}^\dagger c_j \right) - \mu_\mathrm{N} \sum_{j\le L} c_j^\dagger c_j - \mu \sum_{L+1\le j} c_j^\dagger c_j \nonumber\\
                &\quad + \sum_{L+1 \le j} \left( \Delta c_j^\dagger c_{j+1}^\dagger + \mathrm{H.c.} \right)
    +\sum_{1 \le j \le L} V_j c_j^\dagger c_j,
\end{align}
where $V_j \in [-W,W]$ is the random potential at site $j$, $L$ is the length of the diffusive normal metal, and $\mu_\mathrm{N}$ is chemical potential in the normal metal and diffusive normal metal region.

We numerically calculate the spatial dependence of the pair amplitude 
in the topological regime, topological regime near the topological critical point, and non-topological regime. 
In the topological regime, $k_{F} \xi \geq 1$ is satisfied and 
the quasiclassical approximation is applicable. 
In connection with physical observables, we also calculate the local density of states in the diffusive normal metal region and the differential conductance.
These are calculated by the recursive Green's function method \cite{umerski1997closed} {[See \headapp{\ref{sub:recursive}} for details]}.
We take the impurity average of $10^5$ samples.

As a comparison, we also calculate the spatial dependence of the pair amplitude in a normal metal/diffusive normal metal/$s$-wave superconductor junction where the Hamiltonian is obtained from \eq{\ref{eq:hamiltonian-junction}} by replacing $c^\dagger c \to \sum_\sigma c_{\sigma}^\dagger c_{\sigma}$ and $c_j^\dagger c_{j+1}^\dagger\to c_{j,\uparrow}^\dagger c_{j,\downarrow}^\dagger$ with spin index $\sigma$. 
In both the normal metal/diffusive normal metal/Kitaev chain and normal metal/diffusive normal metal/$s$-wave superconductor junctions, 
the superconducting region $(j \geq L+1)$ becomes insulating in the superconducting state for $|\mu| > 2t$. 
Then, it is noted that the topological critical point and metal-insulator transition point coincide in the Kitaev chain region.

We first discuss the spatial dependence of pair amplitudes, which is shown in \fig{\ref{fg:pair-amp}}, in the normal metal/diffusive normal metal/Kitaev chain junction. 
In the topological regime far from the topological critical point,
the odd-frequency spin-triplet even-parity pair amplitude has a sharp peak at the boundary between the diffusive normal metal and Kitaev chain regime {[\fig{\ref{fg:pair-amp}(a)}]}. 
These values are $10^8$ times larger than those at $j=150$. 
The odd-frequency spin-triplet even-parity pair amplitude oscillates in the Kitaev chain region $(j \geq L+1)$.
This oscillation, which is the same effect discussed in the semi-infinite Kitaev chain [\sect{\ref{sect:semi-infinite}}, \fig{\ref{fg:semi-inf}(b)}], comes from Friedel oscillation.
On the other hand, it penetrates the diffusive normal metal without oscillations in the topological regime [\fig{\ref{fg:pair-amp}(a)}]. 

The corresponding spatial dependence of the even-frequency spin-triplet odd-parity pair amplitude is shown in \fig{\ref{fg:pair-amp}(d)}. 
It does not have an oscillatory behavior in the Kitaev chain region and becomes constant far from the boundary. 
Although the even-frequency spin-triplet odd-parity pair amplitude penetrates the diffusive normal metal, its magnitude $0.44$ at $j=25$ is small compared to that of the odd-frequency spin-triplet even-parity pair amplitude which is $19.28$.
Therefore, we conclude that our study indeed verifies the existence of the proximity effect driven by the odd-frequency spin-triplet even-parity pairing in the Kitaev chain junction.

\begin{figure*}[!htbp]
    \centering
    \includegraphics[scale=0.4]{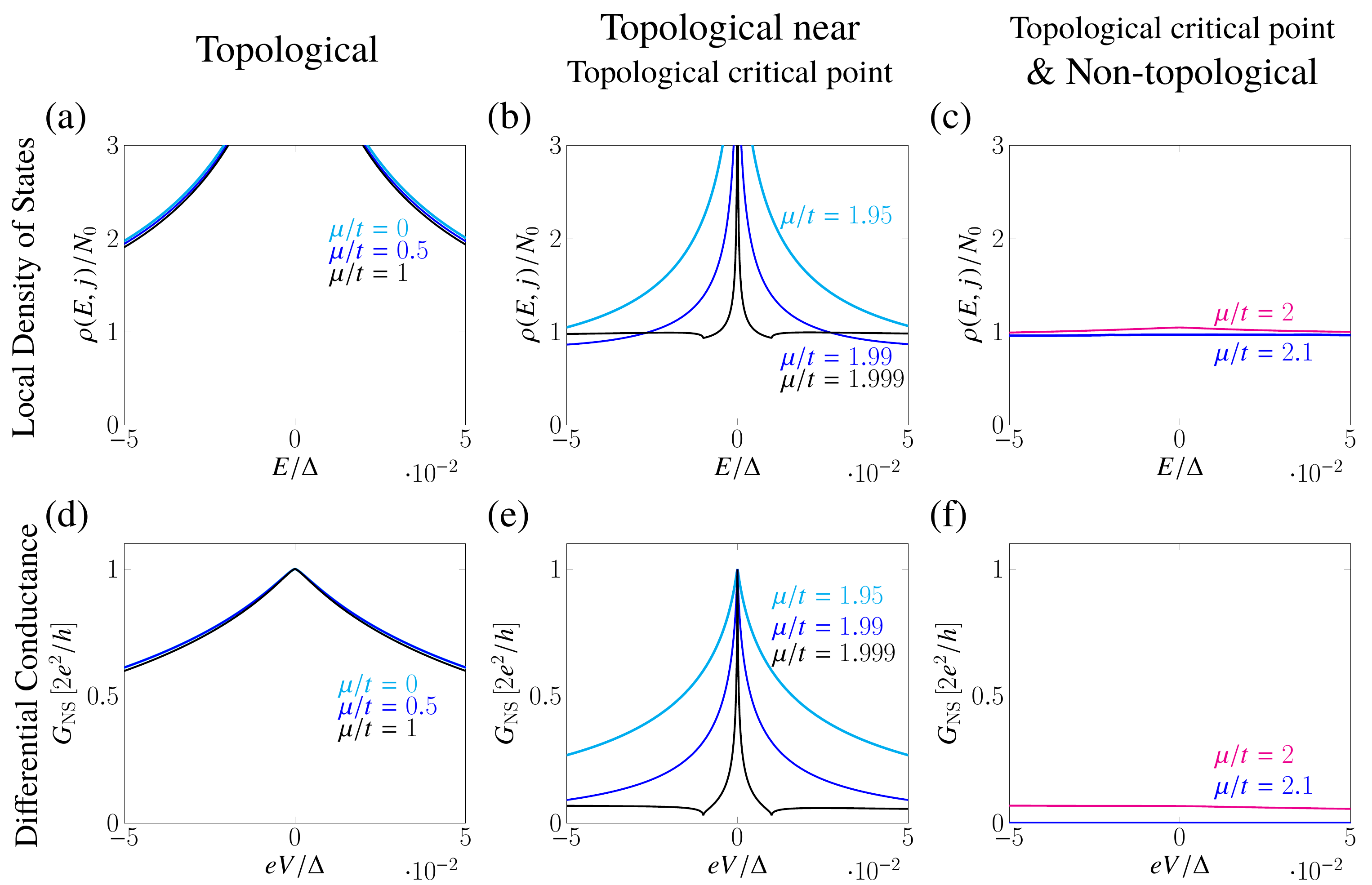}
    \caption{
        (a--c) the local density of states in the diffusive normal metal for energy $E$ in the normal metal/diffusive normal metal/Kitaev chain junction: $j=20$;
        (d--f) differential conductance for bias voltage $eV$ in the normal metal/diffusive normal metal/Kitaev chain junction: $j=-10$;
        $\mu_\mathrm{N}/t=0.5$, $\Delta/t=0.1$, $\omega_n/t=10^{-8}$, $L=20$, and $W/t=1$.
        (a,d) topological regime: $\mu/t=0$, $0.5$, $1$;
        (b,e) topological regime near topological critical point: $\mu/t=1.95$, $1.99$, $1.999$;
        (c,f) topological critical point: $\mu/t=2$; non-topological regime: $\mu=2.1$.
    }
    \label{fg:ldos-cond}
\end{figure*}

Next, we look at the topological regime near the topological critical point. 
As $\mu$ approaches the topological critical point, 
the peak height of the odd-frequency spin-triplet even-parity pair amplitude is suppressed
and the period of its oscillations in the Kitaev chain region is longer
{[\fig{\ref{fg:pair-amp}(b)}]}.
At the topological critical point, the magnitude of the odd-frequency pair amplitude is dramatically suppressed. 
Its peak height is reduced to $1.4\%$ of that for $\mu/t=1.999$. 
However, the proximity effect for the diffusive normal metal still occurs, as seen in \fig{\ref{fg:pair-amp}(c)}
, while the corresponding even-frequency spin-triplet odd-parity pair amplitude hardly penetrates the diffusive normal metal region shown in \fig{\ref{fg:pair-amp}(f)}. 
Also, the odd-frequency spin-triplet even-parity and even-frequency spin-triplet odd-parity pair amplitudes are mixed in the Kitaev chain region at the topological critical point as shown in \fig{\ref{fg:pair-amp}(f)}.

In the non-topological regime, the odd-frequency spin-triplet even-parity pair amplitude 
is dramatically suppressed up to $\sim10^{-3}$. 
On the other hand, although the magnitude of the even-frequency spin-triplet odd-parity pair amplitude is reduced with the increase of $\mu$ in the non-topological (insulating) regime, 
the even-frequency spin-triplet odd-parity pair amplitude does not become zero in the Kitaev chain region shown in \fig{\ref{fg:pair-amp}(f)}. 
Notice that the qualitative behavior of the spatial dependence of the even-frequency spin-triplet odd-parity pairing in the Kitaev chain region, as shown in \figs{\ref{fg:pair-amp}(d--f)}, is insensitive to the change of $\mu$. 

To establish a comparison with a junction coupled to a conventional spin-singlet $s$-wave superconductor, in this part, now we discuss the spatial dependence of the even-frequency spin-singlet even-parity pairing in \figs{\ref{fg:pair-amp}(g--i)}.
When the spin-singlet $s$-wave superconductor is 
in the metallic regime far from the metal-insulator transition point, 
the even-frequency spin-singlet even-parity pair amplitude has a rapid oscillation in the $s$-wave superconductor region and penetrates the diffusive normal metal region without oscillations [\fig{\ref{fg:pair-amp}(g)}]. 
Near the metal-insulator transition point [\fig{\ref{fg:pair-amp}(h)}], the pair amplitude has more oscillations at the diffusive normal metal/Kitaev chain interface while the oscillating period is longer in the $s$-wave superconductor region. 
At the metal-insulator transition point, the oscillatory behavior of the 
spatial dependence of the even-frequency spin-singlet even-parity pair amplitude in the $s$-wave superconductor region
disappears as shown in \fig{\ref{fg:pair-amp}(i)} because $k_F$ becomes zero. 
In the insulating regime, although the magnitude of the even-frequency spin-singlet even-parity pair amplitude in the $s$-wave superconductor region
is suppressed with the increase of $\mu$, 
its value is still nonzero similar to that of 
the even-frequency spin-triplet odd-parity pair amplitude in the normal metal/diffusive normal metal/Kitaev chain junction [\fig{\ref{fg:pair-amp}(i)}]. 
By contrast to the even-frequency spin-triplet odd-parity pair amplitude, the even-frequency spin-singlet even-parity component can penetrate the diffusive normal metal region even in the insulating regime. 

To summarize what we have discussed so far, the spatial dependence of the odd-frequency spin-triplet even-parity pairing is very different from that of the even-frequency spin-triplet odd-parity pair amplitude. 
The odd-frequency spin-triplet even-parity pairing is localized near the boundary between the diffusive normal metal/Kitaev chain in the topological phase 
and its maximum value is suddenly reduced at the topological critical point. 
It almost disappears in the non-topological phase. 
This feature is also very different from the even-frequency spin-singlet even-parity pairing in the normal metal/diffusive normal metal/$s$-wave superconductor junction. 
It is natural to consider that the dramatic suppression of the odd-frequency pair amplitude in the diffusive normal metal at the topological critical point is relevant to the topological transition specific to the Kitaev chain. 
This is because the odd-frequency pairing is enhanced in the topological phase due to the emergence of Majorana fermion which is a particular type of the zero-energy Andreev bound states. 

To establish a relation between odd-frequency pairing and experimental observables, we now investigate the local density of states and differential conductance as shown in \fig{\ref{fg:ldos-cond}}.
In \figs{\ref{fg:ldos-cond}(a--c)} we present the $\rhoj/N_0$ {[\eq{\ref{eq:ldos}}]} in the diffusive normal metal region of the normal metal/diffusive normal metal/Kitaev chain junction, where $N_0$ denotes the density of states at the Fermi level in an infinite ballistic normal metal case without impurity scattering.
In the topological regime {[\fig{\ref{fg:ldos-cond}(a)}]}, 
the local density of states has a sharp zero-energy peak\cite{odd1}.
This peak is correlated with the enhancement of the odd-frequency spin-triplet even-parity shown in \fig{\ref{fg:pair-amp}(a)} due to Majorana fermion.
The peak height and width decrease as $\mu$ approaches the topological critical point {[\fig{\ref{fg:ldos-cond}}(b)]}. 
However, the zero-energy density of states slightly exists at the topological critical point {[\fig{\ref{fg:ldos-cond}(c)}]} 
due to the existence of the odd-frequency pairing shown by a red line in \fig{\ref{fg:pair-amp}(c)}.
This result corresponds to the previous result using the quasiclassical approximation\cite{Higashitani}.
In the non-topological regime, the local density of states has no zero-energy states {[\fig{\ref{fg:ldos-cond}(c)}]}.
This consequence is correlated with the absence of odd-frequency pairing shown by a blue line in \fig{\ref{fg:pair-amp}(c)}.

\begin{figure}[!htbp]
    \centering
    \includegraphics[scale = 0.5]{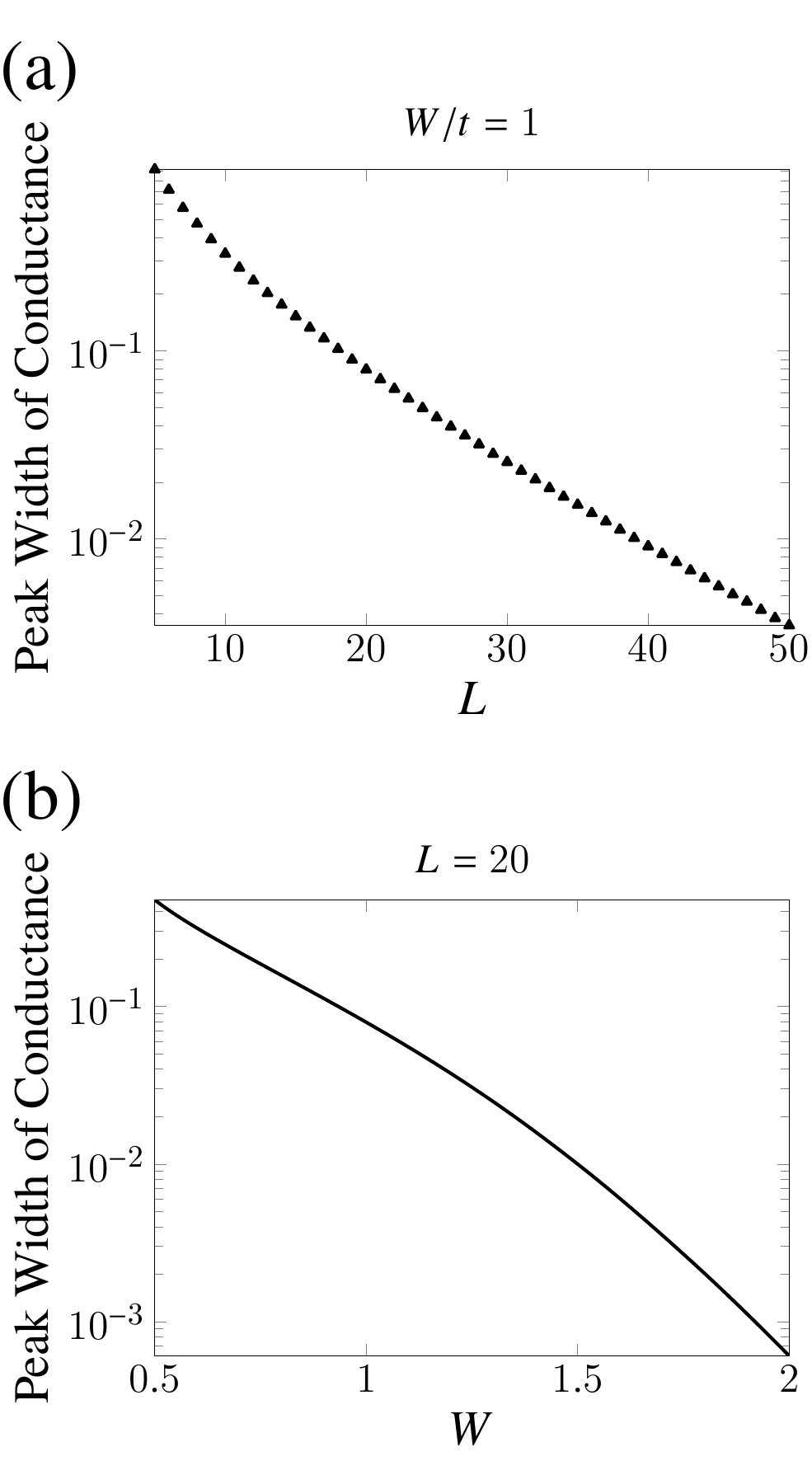}
    \caption{
        Half width at half maximum of the zero-bias conductance peak is plotted as a function of 
        (a) $L$; (b) $W$;
        in the diffusive normal metal in the normal metal/diffusive normal metal/Kitaev chain junction.
        $L$: the length of the diffisive normal metal,
        $W$: the intensity of random potential.
        $\mu_\mathrm{N}/t=0.5$, $\Delta/t=0.1$, $\mu_\mathrm{N}/t=\mu=0$, $j=-10$, and $\delta_\epsilon/t=10^{-8}$.
        (a) $W/t=1$; (b)$L=20$.
    }
    \label{fg:cond-amp}
\end{figure}

\begin{table*}[!htbp]
    \centering
    \caption{
        Anomalous proximity effect to the diffusive normal metal region in the normal metal/diffusive normal metal/Kitaev chain junction. 
        The columns represent the focused regime, the penetration of the odd-frequency spin-triplet even-parity pair amplitude, that of the even-frequency spin-triplet odd-parity pair amplitude, the existence of the zero-energy peak, and the characteristic of the zero-bias conductance peak, respectively. 
        $\bigcirc$ ($\times$) {[$\triangle$]} means that the penetration and spatial oscillation of the pair amplitude, the zero-energy peak, and the zero-bias conductance peak, respectively, (doesn't) {[slightly]} exists. 
    $\circledcirc$ means that the strong resonance occurs at the interface between the diffusive normal metal and Kitaev chain region. 
} 

\scalebox{0.78}{
    \begin{tabular}{c||cc|cc|c|cc}
        \hline
        \multirow{2}{*}{Regime}                                              & \multicolumn{2}{c|}{Odd-frequency spin-triplet even-parity} & \multicolumn{2}{c|}{Even-frequency spin-triplet odd-parity} & Zero-energy peak of  & \multicolumn{2}{c}{Zero-bias conductance peak} \\
                                & Penetration & Spatial oscillation                                                    & Penetration  & Spatial oscillation                                                   & the local density of states                                                                                                            & Quantization & Width         \\\hline\hline
        Topological                                                          & \scalebox{1.5}{$\circledcirc$} & $\times$                                            & $\bigcirc$ & $\times$                                                                & $\bigcirc$                                                                                                          & $\bigcirc$ & wide \\\hline
        Topological near topological critical point & \scalebox{1.5}{$\circledcirc$} & $\bigcirc$                                          & $\bigcirc$ & $\times$                                                                & $\bigcirc$                                                                                                          & $\bigcirc$ & narrow        \\\hline
        Topological critical point                  & $\bigcirc$  & $\bigcirc$                                                             & \scalebox{1.5}{$\triangle$} & $\times$                                               & \scalebox{1.5}{$\triangle$}                                                                                         & \multicolumn{2}{c}{$\times$} \\\hline
        Non-topological                                                      & $\times$  & $\times$                                                                 & $\times$ & $\times$                                                                  & $\times$                                                                                                            & \multicolumn{2}{c}{$\times$} \\\hline
    \end{tabular}
}
    \label{tab:sum-anomalous}
\end{table*}
\begin{table*}[!htbp]
    \centering
    \caption{
        Conventional proximity effect to the diffusive normal metal region in the normal metal/diffusive normal metal/$s$-wave superconductor junction.
        The columns represent the focused regime, the penetration of the even-frequency spin-singlet even-parity pair amplitude, that of the even-frequency spin-triplet odd-parity pair amplitude, the existence of the zero-energy peak, and the characteristic of the zero-bias conductance peak, respectively. 
The usage of the mark is similar to \tab{\ref{tab:sum-anomalous}}.
    }
\scalebox{0.9}{
    \begin{tabular}{c||cc|c|c}
        \hline
        \multirow{2}{*}{Regime} & \multicolumn{2}{c|}{Even-frequency spin-singlet even-parity}                & Zero-energy peak of  & \multirow{2}{*}{Zero-bias conductance peak}  \\
                                & Penetration       & Spatial oscillation &   the local density of states    &                       \\\hline\hline
        Metallic                & $\bigcirc$        & $\times$          & $\times$                     & $\times$ \\\hline
        Metallic near metal-insulator transition point       & $\bigcirc$        & $\bigcirc$          & $\times$                     & $\times$ \\\hline
        Metal-insulator transition point                     & $\bigcirc$        & $\bigcirc$          & $\times$                     & $\times$ \\\hline
        Insulating              & $\bigcirc$        & $\bigcirc$          & $\times$                     & $\times$ \\\hline
    \end{tabular}
}
    \label{tab:sum-conventional}
\end{table*}

\headfigs{\ref{fg:ldos-cond}(d--f)} provides the bias voltage dependence of differential conductance {[\eq{\ref{eq:cond}}]} in the normal metal/diffusive normal metal/Kitaev chain junction;
\fig{\ref{fg:ldos-cond}(d)} shows in the topological regime,
\fig{\ref{fg:ldos-cond}(e)} is in the topological one near the topological critical point,
and \fig{\ref{fg:ldos-cond}(f)} is at the topological critical point and in the non-topological regime.
In the topological regime, 
the conductance has a peak at zero voltage and its value is  always quantized  {[\fig{\ref{fg:ldos-cond}}(e)]}
due to the existence of the zero-energy surface Andreev bound states and chiral symmetry \cite{Ikegaya2016}.
As $\mu$ approaches the topological critical point, the peak width becomes narrower {[\fig{\ref{fg:ldos-cond}}(e)]}.
At the topological critical point, the zero-bias conductance peak is not quantized {[\fig{\ref{fg:ldos-cond}}(f)]} because the gapless state allows a value other than $2e^2/h$ \cite{Ikegaya2016}.
Also, in the non-topological regime, the differential conductance has no value at zero voltage.
To realize the perfect resonance nature at $eV = 0$ more clearly,
we also calculate the standard deviation of the differential conductance in \headapp{\ref{sect:N/DN/KC}}.

The half-width at half maximum of the zero-bias conductance peak is essential for the effect of impurity scattering.
This is because it is roughly consistent with the Thouless energy, which varies with the influence of impurity scattering \cite{Proximityd2}.
To clarify the effect of impurity scattering in the diffusive normal metal,
we calculate the half-width at half maximum of the zero-bias conductance peak in {\fig{\ref{fg:cond-amp}}} {[See \headapp{\ref{sect:hwhm}} for how to calculate the half-width at half maximum].}
\headfig{\ref{fg:cond-amp}(a)} shows the half-width at half maximum as a function of the length L of the diffusive normal metal region;
\fig{\ref{fg:cond-amp}(b)} represents that to the intensity $W$ of the impurity scattering.
These results reveal that the half-width at half maximum of the zero-bias conductance peak is proportional to $\exp(c_1 L + c_2 W)$,
where $c_1$ and $c_2$ are negative constants.
This means that the specificity of zero-bias conductance becomes more pronounced as the impurity scattering increases.
The half-width at half maximum of the zero-energy peak of the local density of states is calculated in \headapp{\ref{sect:N/DN/KC}}. 

We have summarized our obtained results in \tab{\ref{tab:sum-anomalous}}. As a reference, we also show the conventional proximity effect in the normal metal/diffusive normal metal/$s$-wave superconductor junction in \tab{\ref{tab:sum-conventional}}. 

\section{Conclusion}
In this paper, we have studied the spatial dependence of the 
odd-frequency spin-triplet $s$-wave pairing in Kitaev chain junctions. 
First, in the semi-infinite Kitaev chain,  the magnitude of the 
odd-frequency pair amplitude, localized near the edge, decreases towards 
topological critical points. 
Contrary to the common understanding of odd-frequency pairing being generated only at interfaces,
we have found, both numerically and analytically, that the odd-frequency pair amplitude spreads 
into the bulk and takes a constant value at the topological critical point.
In the topological regime, we found that there is a coincidence of 
the spatial dependence of the odd-frequency pair amplitude with that of the local density of states at zero 
energy in a semi-infinite Kitaev chain. 
This result implies that there is a one-to-one correspondence with the wave function of Majorana fermion and the odd-frequency pair amplitude at the low-frequency limit. 

Second, we have also studied the proximity effect in a
normal metal/diffusive normal metal/Kitaev chain junction.  
The magnitude of the odd-frequency pair amplitude 
has a maximum in the Kitaev chain region near the diffusive normal metal/Kitaev chain boundary and 
it decreases towards the topological critical point. 
We have found that the spatial dependence 
of the odd-frequency spin-triplet even-parity pair amplitude is very different from that of the even-frequency spin-triplet odd-parity pair amplitude. 
The odd-frequency spin-triplet even-parity pair amplitude is localized near the boundary between the diffusive normal metal/Kitaev chain 
and suddenly reduced at the topological critical point. 
The dramatic suppression of the 
odd-frequency pair amplitude in the diffusive normal metal
at the topological critical point is relevant to the topological transition since odd-frequency pairing is generated from zero-energy Andreev bound states which is 
nothing but a topological edge state called Majorana fermion. 
Even at the topological critical point, the odd-frequency spin-triplet even-parity pairing causes the proximity effect.
Moreover, the zero-bias conductance is quantized in the topological 
regime and  the local density of states has a zero-energy peak in the diffusive normal metal. 
At the topological critical point, the height of the zero-energy peak of the local density of states is suppressed and 
the zero-bias conductance is no more quantized. 

In conclusion, we have investigated the behavior of the odd-frequency pairing in the Kitaev chain.
Our study would motivate further investigation in more complicated systems containing Rashba nanowires with the magnetic field and $s$-wave superconductivity where Kitaev chain can be realized.

\section*{Acknowledgments}
We thank J.~Cayao for fruitful discussions.
This work was supported by  Grant-in-Aid
for Scientific Research on Innovative Areas, Topological
Material Science (Grant Nos. JP15H05851, 
JP15H05853, and JP15K21717) and Grant-in-Aid for
Scientific Research B (Grant No. JP18H01176) 
from the Ministry of Education, Culture,
Sports, Science, and Technology, Japan (MEXT).

\appendix
\section{General Solution of Majorana Wave Function}
We briefly explain the Majorana wave function [\sect{\ref{sect:semi-infinite}}, \eq{\ref{eq:majowave}}].
We consider the semi-infinite Kitaev chain as shown in \fig{\ref{fg:system}(a)}.
In this system, the analytical solution of a Majorana wave function is derived by Hegde \textit{et al.} \cite{hegde2016majorana}.
The Hamiltonian of this system is written as follows:
\begin{align}
    \label{eq:kitaev-two}
    \mathcal{H} &= - t \sum_{j>0} \left( c_j^\dagger c_{j+1} + c_{j+1}^\dagger c_j \right) - \mu \sum_{j>0} \left(c_j^\dagger c_j-\frac{1}{2}\right) \nonumber\\
                &\quad + \sum_{j>0} \left( \Delta c_j^\dagger c_{j+1}^\dagger + \mathrm{H.c.} \right).
\end{align}
By using the Majorana operators
\begin{align}
    \label{eq:majo-ab}
    \hat{a}_j &= c_j + c_j^\dagger, \quad\hat{b}_j = i(c_j^\dagger - c_j),
\end{align}
the Majorana representation of the Hamiltonian [\eq{\ref{eq:kitaev-two}}]
is obtained as
\begin{align}
    \mathcal{H}_M &= -\frac{i}{2} \sum_{j>0} \left[\left(t-\Delta\right)\ahat_j\bhat_{j+1} - \left(t+\Delta\right)\bhat_j\ahat_{j+1}\right]\nonumber\\
                  &\quad -\frac{i\mu}{2} \sum_{j>0} \ahat_j \bhat_j.
\end{align}
With Heisenberg equation $[\hat{b}_j, \mathcal{H}]=0$,
the site relation of Majorana wave function is given by 
\begin{align}\label{eq:majo}
    \left(t+\Delta\right) a_{j+1} + \left(t-\Delta\right) a_{j-1} + \mu a_j = 0,
\end{align}
where we replace the Majorana wave function $a_j$ for the Majorana operator $\hat{a}_j$.
The application of Z-transform
\begin{align}
    \mathcalz\left[a_j\right] &\equiv \sum_{j=1}^\infty z^{-j} a_j,\\
    \mathcalz\left[a_{j+1}\right] &= z \mathcalz\left[a_j\right] - a_1,\\
    \mathcalz\left[a_{j-1}\right] &= z^{-1} \mathcalz\left[a_j\right],
\end{align}
to \eq{\ref{eq:majo}} gives
\begin{align}
    \mathcalz\left[a_j\right] = \frac{a_1 z^2}{z^2 + \frac{\mu}{t+\Delta}z + \frac{t-\Delta}{t+\Delta}}.
\end{align}
The inverse transform of this provides the Majorana wave function for $0<\Delta<t$
\begin{align}
    a_j &= a_1 C^{j-1} \left\{\cos{\left[\beta (j-1)\right]}+ \frac{1}{\tan{\beta}} \sin{\left[\beta (j-1)\right]}\right\},
\end{align}
where $C = \sqrt{t-\Delta}/\sqrt{t+\Delta}$ and $\beta = \arctan{(\sqrt{4t^2-4\Delta^2-\mu^2}/\mu})$.

\section{Numerical Calculation Method}
\subsection{Recursive Green's function}
\label{sub:recursive}
To obtain the local density of states, differential conductance, and pair amplitude,
we use the recursive Green's function method in \sects{\ref{sect:semi-infinite}} and {\ref{sect:junction}}.
This method is useful in the actual numerical calculation.
The Green's function in a finite system is defined as follows:
\begin{align}
    \label{eq:green-hami}
    G = \left(zI - H\right)^{-1},
\end{align}
where $I$ is an identity matrix with the same size as $H$.
When $N$ denotes the number of sites and $f$ represents
the spin and the electron-hole degree of freedom,
$z$, $H$, and $G$ become $fN\times fN$ matrices.
In the Kitaev chain, $f=2$ holds.
$z$ is defined as
\begin{align}
    z =
    \left\{
        \begin{array}{cl}
            E - i\delta_\epsilon & \mbox{(Advanced Green's function)}\\
            E + i\delta_\epsilon & \mbox{(Retarded Green's function)}\\
                i\omega_n        & \mbox{(Matsubara Green's function)}
        \end{array}
    \right.,
\end{align}
where $E$ stands for energy, $\delta_\epsilon$ denotes an infinitesimal positive number, and $\omega_n$ is a Matsubara frequency.
We express the Hamiltonian as follows:
\begin{align}
    \label{eq:hamiltonian}
    H =
    \begin{bmatrix}
        \check{u}         & \check{t}         &                   &                    & O\\
        \check{t}^\dagger & \check{u}         & \check{t}         &                    & \\
                          & \check{t}^\dagger & \ddots            & \ddots             & \\
                          &                   & \ddots            & \ddots             & \check{t}\\
        O                 &                   &                   & \check{t}^\dagger  & \check{u}
    \end{bmatrix},
\end{align}
where $\check{\hspace{0.5em}}$ denotes a $2\times2$ matrix and
$O$ stands for a zero matrix.
Also, we write the Green's function of the Hamiltonian {[\eq{\ref{eq:hamiltonian}}]} as follows:
\begin{align}
    G =
    \begin{bmatrix}
        \check{G}_{1, 1}^{(N)} & \check{G}_{1, 2}^{(N)} & \cdots & G_{1, N}^{(N)}\\
        \check{G}_{2, 1}^{(N)} & \check{G}_{2, 2}^{(N)} &        & \vdots\\
        \vdots                 &                        & \ddots & \vdots\\
        \check{G}_{N, 1}^{(N)} & \cdots                 & \cdots & \check{G}_{N, N}^{(N)}
    \end{bmatrix},
\end{align}
where the superscript $(N)$ of $\check{G}_{i,j}^{(N)}$ denotes the number of sites in the system.

Between $\check{G}_{i,i}^{(i)}$ and $\check{G}_{i+1,i+1}^{(i+1)}$,
\begin{align}
    \label{eq:green-ite-right}
    \check{G}_{i+1, i+1}^{(i+1)} = \left(\check{z} - \check{u} - \check{t}^\dagger \check{G}_{i, i}^{(i)} \check{t} \right)^{-1},
\end{align}
holds.
With an initial value, $\check{G}_{1,1}^{(1)}=(\check{z}-\check{u})^{-1}$,
$\check{G}_{N,N}^{(N)}$ is obtained by repeating \eq{\ref{eq:green-ite-right}}.
Similarly, 
\begin{align}
    \label{eq:green-ite-left}
    \check{G}_{1, 1}^{(i+1)} &= (\check{z} - \check{u} - \check{t}\check{G}_{1,1}^{(i)}\check{t}^\dagger)^{-1}
\end{align}
is established between $\check{G}_{1,1}^{(i)}$ and $\check{G}_{1,1}^{(i+1)}$.

We describe how to obtain \eq{\ref{eq:green-ite-right}} below.
We divide the Hamiltonian as follows:
\begin{align}
    H =
    \begin{bmatrix}[ccc|c]
        &&&\\
        &H_A&&H_B\\
        &&&\\
        \cmidrule(lr){1-4}
        &H_C&&H_D
    \end{bmatrix},
\end{align}
where
\begin{align}
    H_A &=
    \begin{bmatrix}
        \check{u}         & \check{t}         &           &                   & O\\
        \check{t}^\dagger & \check{u}         & \check{t} &                   & \\
                          & \check{t}^\dagger & \ddots    & \ddots            & \\
                          &                   & \ddots    & \ddots            & \check{t}\\
        O                 &                   &           & \check{t}^\dagger & \check{u}
    \end{bmatrix}, \\
    H_B &=
    \begin{bmatrix}
        \check{O}\\
        \check{O}\\
        \vdots\\
        \check{t}
    \end{bmatrix}, \\
    H_C &=
    \begin{bmatrix}
        \check{O} & \check{O} & \cdots & \check{t}^\dagger
    \end{bmatrix}, \\
    H_D &=
    \begin{bmatrix}
        \check{u}
    \end{bmatrix}.
\end{align}
By using these matrices, the Green's function becomes
\begin{align}
    G &= \left(z I - H\right)^{-1}\nonumber\\
      &=
    \begin{bmatrix}[ccc|c]
        &                                &&\\
        & z I_A - H_A && -H_B \\
        &                                &&\\
        \cmidrule(lr){1-4}
        & -H_C                           && z I_D - H_D
    \end{bmatrix}^{-1},
\end{align}
where $I_A$ and $I_D$ are identity matrices with the same size as $H_A$ and
$H_D$, respectively.
The relation of the inverse matrix for the block matrix
provides
\begin{align}
    G &=
    \begin{bmatrix}[ccc|c]
        &         &&\\
        & A && B \\
        &         &&\\
        \cmidrule(lr){1-4}
        & C    && D
    \end{bmatrix}^{-1}\nonumber\\
    &=
    \begin{bmatrix}[ccc|c]
        &         &&\\
        & A^{-1}+A^{-1}BS^{-1}CA^{-1} && -A^{-1}BS^{-1} \\
        &         &&\\
        \cmidrule(lr){1-4}
        & -S^{-1}CA^{-1}    && S^{-1}
    \end{bmatrix}\nonumber\\
    &=
    \begin{bmatrix}[ccc|c]
        \check{G}_{1, 1}^{(N)} & \check{G}_{1, 2}^{(N)} & \cdots & \check{G}_{1, N}^{(N)}\\
        \check{G}_{2, 1}^{(N)} & \check{G}_{2, 2}^{(N)} &        & \vdots\\
        \vdots                 &                        & \ddots & \vdots\\
        \cmidrule(lr){1-4}
        \check{G}_{N, 1}^{(N)} & \cdots                 & \cdots & \check{G}_{N, N}^{(N)}
    \end{bmatrix}.
\end{align}
where $A = zI_A - H_A$, $B = -H_B$, $C = -H_C$, $D = zI_D - H_D$, and $S=D-CA^{-1}B$.
Thus, the recurrence relation \eq{\ref{eq:green-ite-right}} is
transformed as follows:
\begin{align}
    \check{G}_{N, N}^{(N)} &= \left(D - CA^{-1}B\right)^{-1}\nonumber\\
    \label{eq:recurrence-formula}
                           &= \left(\check{z} - \check{u} - \check{t}^\dagger \check{G}_{N-1, N-1}^{(N-1)}\check{t}\right)^{-1}.
\end{align}

We consider connecting $\check{G}_{M,M}^{(M)}$ and $\check{G}_{1,1}^{(N)}$.
The connected Green's function is written as 
\begin{align}
    G =
    \begin{bmatrix}
        \ddots &  &  &\\
               & \check{G}_{M, M  }^{(M+N)} & \check{G}_{M,   M+1}^{(M+N)} &\\
               & \check{G}_{M+1, M}^{(M+N)} & \check{G}_{M+1, M+1}^{(M+N)} &\\
               & & & \ddots
    \end{bmatrix}.
\end{align}
Then, the required local Green's functions are obtained as follows:
\begin{align}
    \check{G}_{M, M    }^{(M+N)} &= \left[\left(\check{G}_{M,M}^{(M)}\right)^{-1} - \check{t} \check{G}_{1,1}^{(N)}\check{t}^\dagger\right]^{-1},
    \label{eq:green-connect-one}\\
    \check{G}_{M+1, M+1}^{(M+N)} &= \left[\left(\check{G}_{1,1}^{(N)}\right)^{-1} - \check{t}^\dagger \check{G}_{M,M}^{(M)} \check{t}\right]^{-1},
    \label{eq:green-connect-two}\\
    \check{G}_{M, M+1  }^{(M+N)} &= \check{G}_{M,M}^{(M+N)} \check{t} \check{G}_{1,1}^{(N)},
    \label{eq:green-connect-three}\\
    \check{G}_{M+1, M  }^{(M+N)} &= \check{G}_{1,1}^{(N)} \check{t}^\dagger \check{G}_{M,M}^{(M+N)}.
    \label{eq:green-connect-four}
\end{align}
where $\check{t}$ is a hopping matrix between connected two sites.

To obtain the Green's function at the right end of the semi-infinite system, 
we briefly explain the M\"{o}bius transformation.
There are right-hand and left-hand M\"{o}bius transformations.
Here, we consider only the left-hand transformation.
We define the left-hand transformation as 
\begin{align}
    \label{eq:mobius}
    \begin{bmatrix}
        A & B\\
        C & D
    \end{bmatrix}
    \mebius Y \equiv (AY+B)(CY+D)^{-1},
\end{align}
where $A$, $B$, $C$, $D$, and $Y$ are $N\times N$ matrices.
The following coupling law 
\begin{align}
    \label{eq:mobius-connect}
    E \mebius \left(F \mebius Y\right) = (EF) \mebius Y,
\end{align}
holds for M\"{o}bius transformation, 
where $E$ and $F$ are $2N\times2N$ matrices.

We express the relation between $\check{G}_{N,N}^{(N)}$ and $\check{G}_{N-1,N-1}^{(N-1)}$ by M\"{o}bius transformation as follows:
\begin{align}
    \check{G}_{N, N}^{(N)} &= \left(\check{z} - \check{u} - \check{t}^\dagger \check{G}_{N-1, N-1}^{(N-1)} \check{t} \right)^{-1}\nonumber\\
                           &= \check{t}^{-1} \left[(\check{z} - \check{u}) \check{t}^{-1} - \check{t}^\dagger \check{G}_{N-1, N-1}^{(N-1)}\right]^{-1}\nonumber\\
    \label{eq:mobius-recurrence}
                           &=
    \begin{bmatrix}
        \check{O}          & \check{t}^{-1}\\
        -\check{t}^\dagger & (\check{z} - \check{u})\check{t}^{-1}
    \end{bmatrix}
    \mebius \check{G}_{N-1, N-1}^{(N-1)},
\end{align}
where we choose $A=\check{O}$, $B=\check{t}^{-1}$, $C=-\check{t}^\dagger$, $D=(\check{z}-\check{u})\check{t}^{-1}$, and $Y=\check{G}_{N-1,N-1}^{(N-1)}$ in \eq{\ref{eq:mobius}}.
We define the left part of the M\"{o}bius transformation in \eq{\ref{eq:mobius-recurrence}} as 
\begin{align}
    \label{eq:matrixx}
    X =
    \begin{bmatrix}
        \check{O}          & \check{t}^{-1}\\
        -\check{t}^\dagger & (\check{z} - \check{u})\check{t}^{-1}
    \end{bmatrix}.
\end{align}
By repeating to fix the form of M\"{o}bius transformation in \eq{\ref{eq:mobius-recurrence}} and using 
\eq{\ref{eq:mobius-connect}}, $\check{G}_{N, N}^{(N)}$ is transformed as 
\begin{align}
    \label{eq:mobius-ite}
    \check{G}_{N, N}^{(N)} &= X \mebius \check{G}_{N-1, N-1}^{(N-1)}\nonumber\\
                           &= X \mebius (X \mebius \check{G}_{N-2, N-2}^{(N-2)})\nonumber\\
                           &\quad\vdots\nonumber\\
                           &= X^{N-1} \mebius \check{G}_{1, 1}^{(1)}.
\end{align}
The eigenvalue decomposition of $X$ is given as follows:
\begin{align}
    \label{eq:x-eig}
    X &= Q
    \begin{bmatrix}
        \check{\Lambda}_1 & \\
                          & \check{\Lambda}_2
    \end{bmatrix}
    Q^{-1},\\
    \check{\Lambda}_1 &=
    \begin{bmatrix}
        \lambda_1 & \\
                  & \lambda_2
    \end{bmatrix},\quad
    \check{\Lambda}_2 =
    \begin{bmatrix}
        \lambda_3 & \\
                  & \lambda_4
    \end{bmatrix},
\end{align}
where the eigenvalues $\lambda_1$, $\lambda_2$, $\lambda_3$, and $\lambda_4$ of $X$
satisfy $|\lambda_1|< |\lambda_2| < |\lambda_3| < |\lambda_4|$.
Here, it is noted that
degeneracies of eigenvalues are removed by introducing the
infinitesimal imaginary number of the energy of Green's function.
By substituting \eq{\ref{eq:x-eig}} for \eq{\ref{eq:mobius-ite}} and
using the coupling law \eq{\ref{eq:mobius-connect}},
$\check{G}_{N, N}^{(N)}$ becomes
\begin{align}
    \check{G}_{N, N}^{(N)} &= Q
    \begin{bmatrix}
        \check{\Lambda}_1^{N-1} & \\
                                & \check{\Lambda}_2^{N-1}
    \end{bmatrix}
    Q^{-1} \mebius \check{G}_{1, 1}^{(1)}\nonumber\\
    &= Q \mebius \left\{
    \begin{bmatrix}
        \check{\Lambda}_1^{N-1} & \\
                                & \check{\Lambda}_2^{N-1}
    \end{bmatrix}
\mebius \left( Q^{-1} \mebius \check{G}_{1, 1}^{(1)} \right) \right\}.
\end{align}
By performing M\"{o}bius transformation with
$A = \check{\Lambda}_1^{N-1}$, $B=\check{O}$, $C=\check{O}$, $D=\check{\Lambda}_2^{N-1}$,
$Y=Q^{-1}\mebius\check{G}_{1,1}^{(1)}$, $\check{G}_{N, N}^{(N)}$ is transformed as 
\begin{align}
    \label{eq:q-lam-y-lam}
    \check{G}_{N, N}^{(N)} &= Q \mebius \left[\check{\Lambda}_1^{N-1} Y \left(\check{\Lambda}_2^{N-1}\right)^{-1}\right].
\end{align}
For a general $2\times2$ matrix
\begin{align}
    Y &=
    \begin{bmatrix}
        a & b\\
        c & d
    \end{bmatrix},
\end{align}
\begin{align}
    \label{eq:lam-y-lam}
    \lim_{N \to \infty} \check{\Lambda}_1^{N-1} Y \left(\check{\Lambda}_2^{N-1}\right)^{-1} &=
    \lim_{N \to \infty}
    \begin{bmatrix}
        \left(\frac{\lambda_1}{\lambda_2}\right)^{N-1} a & \left(\frac{\lambda_1}{\lambda_4}\right)^{N-1} b\\
        \left(\frac{\lambda_2}{\lambda_3}\right)^{N-1} c & \left(\frac{\lambda_2}{\lambda_4}\right)^{N-1} d
    \end{bmatrix}\nonumber \\
    &= \check{O}
\end{align}
can be confirmed.
Taking the limit $N\to\infty$ in \eq{\ref{eq:q-lam-y-lam}},
we obtain the Green's function $\check{G}_{L}^\infty$ of the right end on the left semi-infinite system as follows:
\begin{align}
    \label{eq:green-left}
    \check{G}_{L}^\infty &= Q \mebius \check{O}\nonumber\\
                         &= \check{Q}_{12} \check{Q}_{22}^{-1}.
\end{align}
The Green's function $\check{G}_R^{\infty}$ is obtained
by exchanging $\check{t}\leftrightarrow t^\dagger$ in $X$ given in \eq{\ref{eq:matrixx}}.
Combining \eqs{\ref{eq:green-ite-right}}, (\ref{eq:green-ite-left}), (\ref{eq:green-connect-one})-(\ref{eq:green-connect-four}),
and (\ref{eq:green-left}),
we can get Green's function of a required site.

\subsection{Calculation of peak width}
\label{sect:hwhm}
\begin{figure}[!htbp]
    \centering
    \includegraphics[scale=0.5]{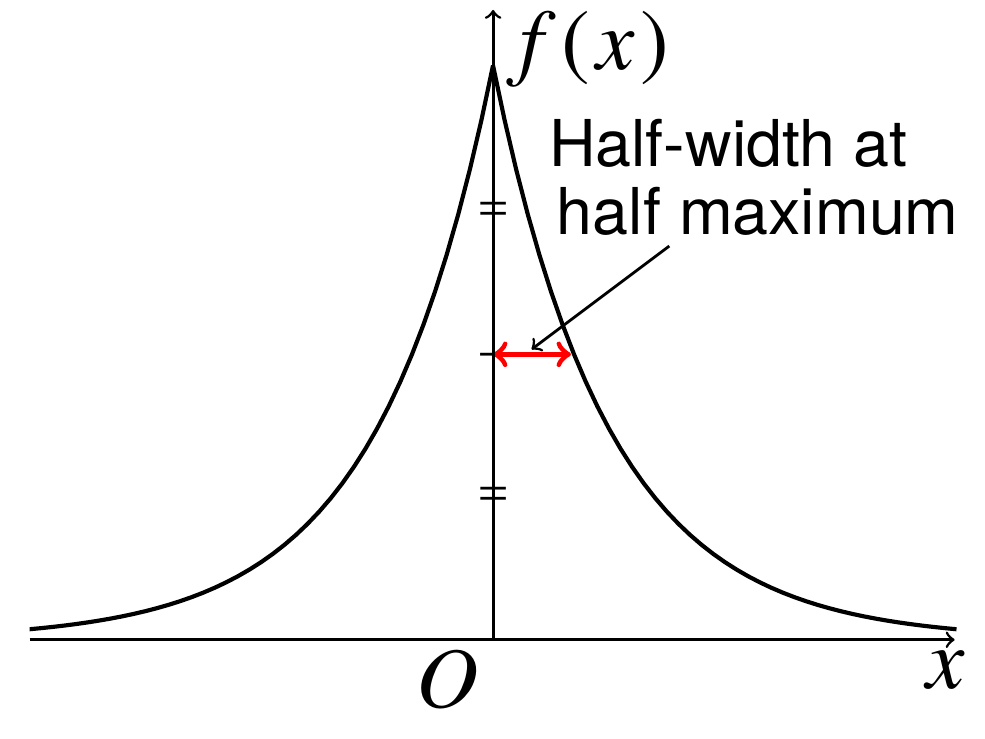}
    \caption{
        The definition of the peak width.
        In this paper, the peak width is defined as the half-width at half maximum.
    }
    \label{fg:hwhm}
\end{figure}
\begin{figure*}[!htbp]
    \centering
    \includegraphics[scale=0.4]{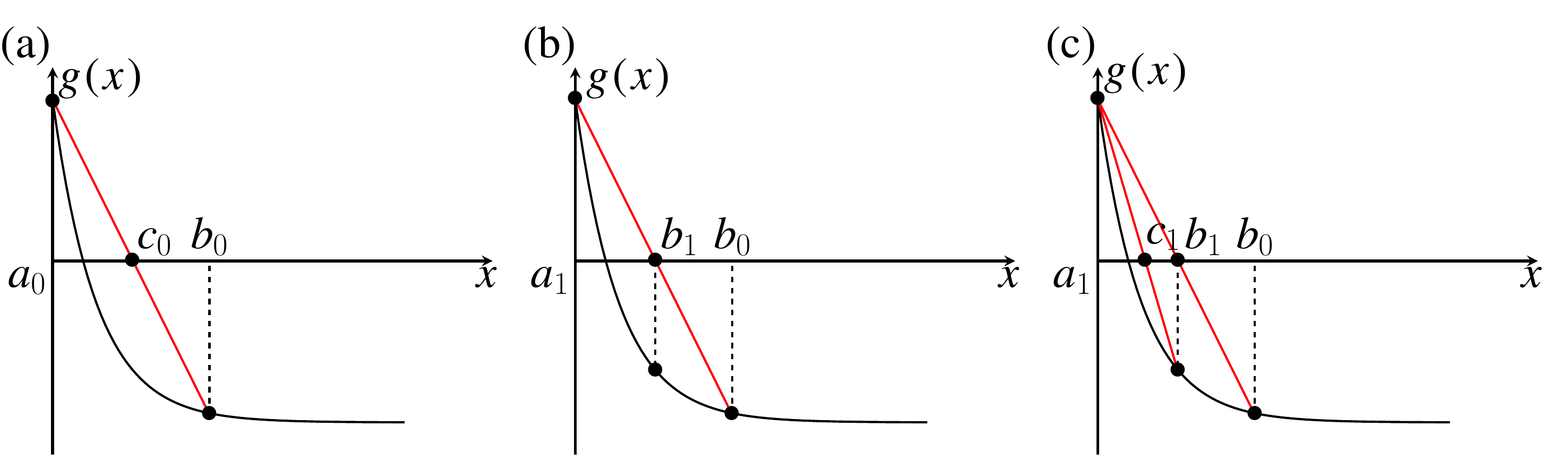}
    \caption{
        Schematics for obtaining the half-width at half maximum of the peak using the false position method.
    }
    \label{fg:f-p-method}
\end{figure*}

We describe how to calculate the peak width shown in \sect{\ref{sect:junction}} \fig{\ref{fg:cond-amp}}.
\headfig{\ref{fg:hwhm}} shows the definition of the peak width as half-width at half maximum.
It is now possible to calculate the output $f(x)$ from the input $x$ by using the recursive Green's function method, where
\begin{align}
    x &=
    \left\{
        \begin{array}{cl}
            E/\Delta & \mbox{(for the local density of states)}\\
            eV/\Delta & \mbox{(for conductance)}
        \end{array}
    \right. ,\\
    f(x) &=
    \left\{
        \begin{array}{cl}
            \rho(E,j)/N_0      & \mbox{(for the local density of states)}\\
            G_\mathrm{NS}[2e^2/h] & \mbox{(for conductance)}
        \end{array}
    \right. .
\end{align}
However, it is difficult to know the analytical functional form of $f(x)$.
Then, we calculate the half-width at half maximum numerically.
We find a solution that satisfies
\begin{align}
    f(x) = \frac{1}{2}f(0).
\end{align}
We define a new function $g(x)$ that shifts $f(x)$ by half of the peak height as
\begin{align}
    g(x) \equiv f(x) - \frac{1}{2}f(0).
\end{align}
The solution of $g(x)=0$ is the half-width at half maximum and is calculated by using the false position method as shown in \headalg{\ref{alg:f-p-method}}.
\begin{algorithm}[H]
    \caption{False position method}
    \label{alg:f-p-method}
    \begin{algorithmic}[1]
        \State{Initial value: $a_0$,$b_0$, Solution: $c \in [a_0, b_0]$}
        \For{$k = 0, 1, 2, \dots$}
            \State{$c_k = \frac{a_kg(b_k) - b_kg(a_k)}{g(b_k) - g(a_k)}$;}
            \If{$g(b_k) g(c_k) < 0$}
                \State{$a_{k+1} = c_k$;}
                \State{$b_{k+1} = b_k$;}
            \Else
                \State{$b_{k+1} = c_k$;}
                \State{$a_{k+1} = a_k$;}
            \EndIf
            \If{$|g(c_k)| < \delta$ \textbf{or} $|a_{k+1} - b_{k+1}| < \epsilon$}
                \State{\textbf{break}}
                \State{$c_k \to c$;}
            \EndIf
        \EndFor
    \end{algorithmic}
\end{algorithm}

We briefly explain the false position method.
First, initial values $a_0$, $b_0$ are determined so that only one solution exists in $[a_0, b_0]$.
Second, let $c_0$ be the intersection of the straight line passing through the points $(a_0, g(0))$ and $(b_0, g(b_0))$ with the $x$-axis [\fig{\ref{fg:f-p-method}(a)}].
When $g(a_0)g(c_0)<0$, there is a solution in $[a_0, c_0]$ and then we set $a_1=a_0$, $b_1=c_0$ [corresponding to \fig{\ref{fg:f-p-method}}(b)].
When $g(c_0)g(b_0)<0$, on the other hand, there is a solution in $[c_0, b_0]$ and then we set $a_1=c_0$, $b_1=b_0$.
Third, the intersection of the straight line, passing through the points $(a_1, g(a_1))$ and $(b_1, g(b_1))$ with $x$ axis, is defined as $c_1$ [\fig{\ref{fg:f-p-method}}(c)]. 
Finally, we obtain the solution by repeating the same operation.
In actual numerical calculations, we set initial values: $a_0=0$, $b_0=1$, and parameters for convergence determination: $\delta=10^{-12}$, $\epsilon=10^{-14}$.

\section{Proof of $-\mathrm{Im}[G_{1,1}] = \mathrm{Im}[\tilde{F}_{1,1}]$ for $\Delta=t$ in Semi-infinite Kitaev Chain}
\label{sec:proof-delta-t}
In this section, we prove \eq{\ref{eq:majo-odd-ana}}
\begin{align}
    -\mathrm{Im}\left[G_{1,1}\right] = \mathrm{Im}\left[\tilde{F}_{1,1}\right],\nonumber
\end{align}
for $\Delta=t$ in \sect{\ref{sect:semi-infinite}}.
We rewrite \eq{\ref{eq:green-ite-left}}
\begin{align}
    \label{eq:semi-inf-green-ite}
    \check{G}_{i+1}^{(i+1)} = \left(\check{z} - \check{u} - \check{t} \check{G}_{i}^{(i)} \check{t}^\dagger \right)^{-1},
\end{align}
where
\begin{align}
    \check{z} &=
    \begin{bmatrix}
        i\omega_n & 0\\
        0         & i\omega_n
    \end{bmatrix},\quad
    \check{u} =
    \begin{bmatrix}
        -\mu & 0\\
        0    & \mu
    \end{bmatrix},\\
    \check{t} &=
    \begin{bmatrix}
        -t      & \Delta\\
        -\Delta & t
    \end{bmatrix}
    = t
    \begin{bmatrix}
        -1 & 1\\
        -1 & 1
    \end{bmatrix}.
\end{align}
We express $\check{G}_{1,1}^{(i)}$ as $\check{G}_{i}$ 
and $\check{v}$ as $\check{z}-\check{u}$ below.
Since $\check{t}$ is a singular matrix, we cannot calculate the leftmost Green's function of the semi-infinite system by simple diagonalization of $X$ in \eq{\ref{eq:matrixx}}.
Then, we first calculate $\check{G}_{i+1}$ with \eq{\ref{eq:semi-inf-green-ite}}:
\begin{align}
    \check{G}_{i+1} &= \left(\check{v}-\check{t}\check{G}_i\check{t}^\dagger\right)^{-1}\nonumber\\
                    &=
    \begin{bmatrix}
        v_{11} - g_i & -g_i\\
        -g_i         & v_{22} - g_i
    \end{bmatrix}^{-1}\nonumber\\
    \label{eq:semi-inf-green-ione}
                    &= \frac{1}{v_{11}v_{22}-(v_{11}+v_{22})g_i}
    \begin{bmatrix}
        v_{22}-g_i & g_i\\
        g_i        & v_{11}-g_i
    \end{bmatrix},
\end{align}
where $g_i$ is defined as
\begin{align}
    \label{eq:green-sum}
    g_i \equiv t^2\left[\left(\check{G}_i\right)_{11}-\left(\check{G}_i\right)_{12}-\left(\check{G}_i\right)_{21}+\left(\check{G}_i\right)_{22}\right].
\end{align}
\headeqs{\ref{eq:semi-inf-green-ione}} and (\ref{eq:green-sum}) provide the relation:
\begin{align}
    \frac{g_{i+1}}{t^2}
    &= \frac{v_{11} + v_{22} - 4g_i}{v_{11}v_{22} - \left(v_{11}+v_{22}\right)g_i}\nonumber\\
    \label{eq:semi-inf-green-sum-ione}
    &= \frac{1}{v_{11}+v_{22}}\left\{4+\frac{\left(v_{11}-v_{22}\right)^2}{v_{11}v_{22}-\left(v_{11}+v_{22}\right)g_i}\right\}.
\end{align}
Let us consider separately for $\mu=0$ and $\mu\neq0$ below.

For $\mu=0$, the second term in \eq{\ref{eq:semi-inf-green-sum-ione}} disappears and $g_i$ becomes
\begin{align}
    g_i = g_\infty &= \frac{2t^2}{i\omega_n}.
\end{align}
By taking the limit $i\to\infty$, \eq{\ref{eq:semi-inf-green-ione}} is transformed as follows:
\begin{align}
    \check{G}_{\infty} &= \frac{1}{z_{11}z_{22}-\left(z_{11}+z_{22}\right)g_\infty}
    \begin{bmatrix}
        z_{22} - g_\infty & g_\infty\\
        g_\infty          & z_{11} - g_\infty
    \end{bmatrix}\nonumber\\
    \label{eq:mu-eq-zero-ginf}
    &= \frac{1}{i\omega_n(\omega_n^2+4t^2)}
    \begin{bmatrix}
        \omega_n^2+2t^2 & -2t^2\\
        -2t^2 & \omega_n^2+2t^2
    \end{bmatrix}.
\end{align}
For $\omega_n\to0$, $-\mathrm{Im}[(\check{G}_\infty)_{11}]\to\infty$ and
$\mathrm{Im}[(\check{G}_\infty)_{21}]\to\infty$ are confirmed; the former means the existence of Majorana fermion.
From \eq{\ref{eq:mu-eq-zero-ginf}}, we obtain
\begin{align}
    &\lim_{\omega_n \to 0} \left\{\mathrm{Im}\left[G_{1,1}\right] + \mathrm{Im}\left[\tilde{F}_{1,1}\right]\right\}\nonumber\\
    &\quad= \lim_{\omega_n\to0} \left\{\mathrm{Im}\left[\left(\check{G}_\infty\right)_{11}\right] + \mathrm{Im}\left[\left(\check{G}_\infty\right)_{21}\right]\right\}\nonumber\\
    &\quad= \lim_{\omega_n\to0} \frac{-\omega_n}{\omega_n^2+4t^2}\nonumber\\
    &\quad= 0,\nonumber\\
    &\quad\Leftrightarrow-\lim_{\omega_n \to 0} \left\{\mathrm{Im}\left[G_{1,1}\right]\right\}=\lim_{\omega_n \to 0} \left\{\mathrm{Im}\left[\tilde{F}_{1,1}\right]\right\}.
\end{align}

For $\mu\neq0$, denoting $\alpha$ and $\beta$ as
\begin{align}
    \alpha &= v_{11}+v_{22}\nonumber\\
    \label{eq:alpha}
           &= 2i\omega_n,\\
    \beta  &= v_{11}v_{22}\nonumber\\
    \label{eq:beta}
           &= -\omega_n^2 -\mu^2.
\end{align}
Moreover, taking the limit $i\to\infty$ in \eq{\ref{eq:semi-inf-green-sum-ione}},
we obtain $g_\infty$ as follows:
\begin{align}
    &\frac{g_\infty}{t^2} = \frac{\alpha-4g_\infty}{\beta-\alpha g_\infty},\nonumber\\
    \label{eq:ginf}
    &\quad\Leftrightarrow g_\infty = \frac{\beta+4t^2\pm\sqrt{\left(\beta+4t^2\right)^2-4\alpha^2t^2}}{2\alpha}.
\end{align}
From \eq{\ref{eq:semi-inf-green-ione}}, $\check{G}_\infty$ becomes
\begin{align}
    \check{G}_\infty &= \frac{1}{-\omega_n^2-\mu^2-\alpha g_\infty}\nonumber\\
    \label{eq:mu-neq-zero-ginf}
                     &\quad\times
    \begin{bmatrix}
        i\omega_n-\mu-g_\infty & g_\infty\\
        g_\infty        & i\omega_n+\mu-g_\infty
    \end{bmatrix}.
\end{align}
We take the plus sign in \eq{\ref{eq:ginf}}.
From \eqs{\ref{eq:alpha}}--(\ref{eq:ginf}),
$\lim_{\omega_n\to0}\alpha g_\infty$ and $\lim_{\omega_n\to0} \mathrm{Im} \left[g_\infty\right]$
are given as
\begin{align}
    \lim_{\omega_n\to0}\alpha g_\infty & = \lim_{\omega_n\to0} \frac{1}{2}\left\{-\omega_n^2-\mu^2+4t^2\right.\nonumber\\
                                       & \quad\left.+\sqrt{\left(-\omega_n^2-\mu^2+4t^2\right)^2+16\omega_n^2t^2}\right\}\nonumber\\
    \label{eq:alphaginf}
                                       & =  
    \left\{
        \begin{array}{cl}
            -\mu^2+4t^2 & (\mbox{Topological})\\
            0           & (\mbox{Non-topological})
        \end{array}
    \right.,\\
    \lim_{\omega_n\to0} \mathrm{Im} \left[g_\infty\right]
    &= \lim_{\omega_n \to 0} -\frac{1}{4} \left[-\omega_n - \frac{\mu^2-4t^2}{\omega_n}\right.\nonumber\\
    &\quad \left.+ \sqrt{\left(-\omega_n - \frac{\mu^2-4t^2}{\omega_n}\right)^2 + 16t^2}\right]\nonumber\\
    &\simeq \lim_{\omega_n \to 0} \frac{1}{4} \left(\frac{\mu^2-4t^2}{\omega_n} - \frac{|\mu^2-4t^2|}{\omega_n}\right)\nonumber\\
    \label{eq:lim-ginf-nontopo}
    &= \left\{
    \begin{array}{cl}
        -\infty & (\mbox{Topological})\\
        0      & (\mbox{Non-topological})
    \end{array}\right..
\end{align}
With \eqs{\ref{eq:mu-neq-zero-ginf}}--(\ref{eq:lim-ginf-nontopo}),
\begin{align}
    -\lim_{\omega_n \to 0} \mathrm{Im}\left[\left(\check{G}_\infty\right)_{11}\right]
    &= \lim_{\omega_n \to 0} \frac{\omega_n - \mathrm{Im}\left[g_\infty\right]}{\omega_n^2+\mu^2+\alpha g_\infty}\nonumber\\
    \label{eq:lim-ginfoneone-two}
    &= \left\{
    \begin{array}{cl}
        \infty & (\mbox{Topological})\\
        0      & (\mbox{Non-topological})
    \end{array}\right.,\\
    \lim_{\omega_n \to 0} \mathrm{Im}\left[\left(\check{G}_\infty\right)_{21}\right]
    &= \lim_{\omega_n \to 0} \frac{\mathrm{Im}\left[g_\infty\right]}{-\omega_n^2-\mu^2-\alpha g_\infty}\nonumber\\
    \label{eq:lim-ginftwoone-two}
    &= \left\{
    \begin{array}{cl}
        \infty & (\mbox{Topological})\\
        0      & (\mbox{Non-topological})
    \end{array}\right.
\end{align}
can be confirmed.
\headeq{\ref{eq:lim-ginfoneone-two}} is consistent with the presence of Majorana fermion
in the topological regime and with the absence of that in the non-topological regime.
As it is for the plus sign, we take the minus sign in \eq{\ref{eq:ginf}}.
The results show
\begin{align}
    \label{eq:lim-ginfoneone-two-minus}
    -\lim_{\omega_n \to 0} \mathrm{Im}\left[\left(\check{G}_\infty\right)_{11}\right]
    &= \left\{
    \begin{array}{cl}
        0           & (\mbox{Topological})\\
        \infty      & (\mbox{Non-topological})
    \end{array}\right.,\\
    \lim_{\omega_n \to 0} \mathrm{Im}\left[\left(\check{G}_\infty\right)_{21}\right]
    &= \left\{
    \begin{array}{cl}
        0      & (\mbox{Topological})\\
        \infty & (\mbox{Non-topological})
    \end{array}\right..
\end{align}
\headeq{\ref{eq:lim-ginfoneone-two-minus}} contradicts the presence of Majorana fermion
in the topological regime.
Then, it is reasonable to take the plus sign in \eq{\ref{eq:ginf}}.
From \eqs{\ref{eq:alphaginf}}, (\ref{eq:lim-ginfoneone-two}), and (\ref{eq:lim-ginftwoone-two}),
we obtain
\begin{align}
    &\lim_{\omega_n \to 0} \left\{\mathrm{Im}\left[G_{1,1}\right] + \mathrm{Im}\left[\tilde{F}_{1,1}\right]\right\}\nonumber\\
    &\quad= \lim_{\omega_n\to0} \left\{\mathrm{Im}\left[\left(\check{G}_\infty\right)_{11}\right] + \mathrm{Im}\left[\left(\check{G}_\infty\right)_{21}\right]\right\}\nonumber\\
    &\quad= \lim_{\omega_n\to0}
    \left\{
        \begin{array}{cl}
             \frac{-\omega_n}{\omega_n^2+4t^2} & (\mbox{Topological})\\
             \frac{-\omega_n}{\omega_n^2+\mu^2} & (\mbox{Non-topological})\\
        \end{array}
    \right.
    \nonumber\\
    &\quad= 0,\\
    &\quad\Leftrightarrow-\lim_{\omega_n \to 0} \left\{\mathrm{Im}\left[G_{1,1}\right]\right\}=\lim_{\omega_n \to 0} \left\{\mathrm{Im}\left[\tilde{F}_{1,1}\right]\right\}.
\end{align}

\begin{figure}[!htbp]
    \centering
    \includegraphics[scale=0.5]{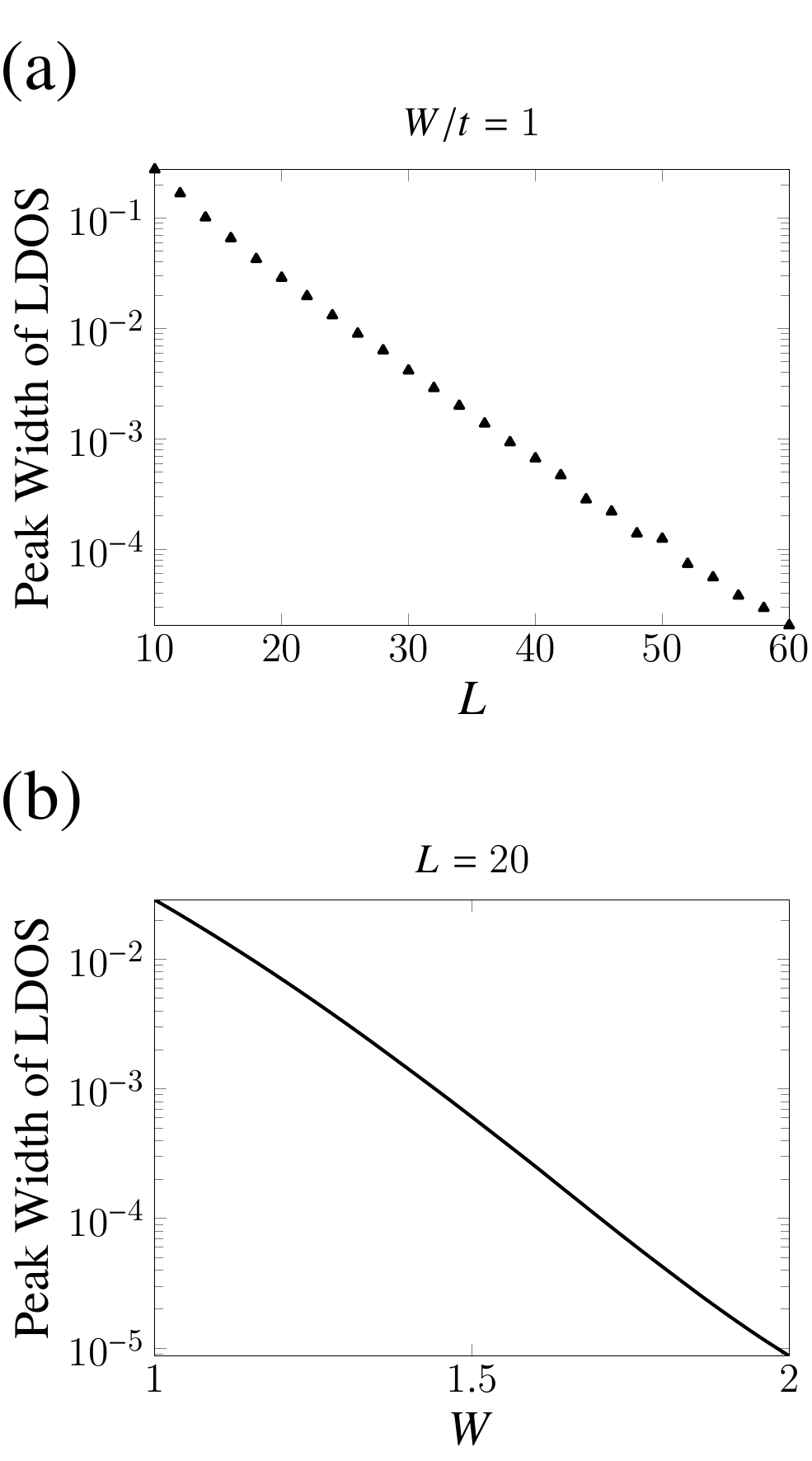}
    \caption{
        Half-width at half maximum of the zero-energy peak of the local density of states is plotted as a function of 
        (a) $L$; (b) $W$;
        of the diffusive normal metal in the normal metal/diffusive normal metal/Kitaev chain junction.
        $L$: the length of the diffusive normal metal,
        $W$: the magnitude of random potential.
        $\Delta/t=0.1$, $\mu_\mathrm{N}=\mu=0$, $j=L/2$, and $\delta_\epsilon/t=10^{-8}$.
        (a) $W=1$; (b)$L=20$.
    }
    \label{fg:ldos-amp}
\end{figure}

\begin{figure}[!htbp]
    \centering
    \includegraphics[scale = 0.4]{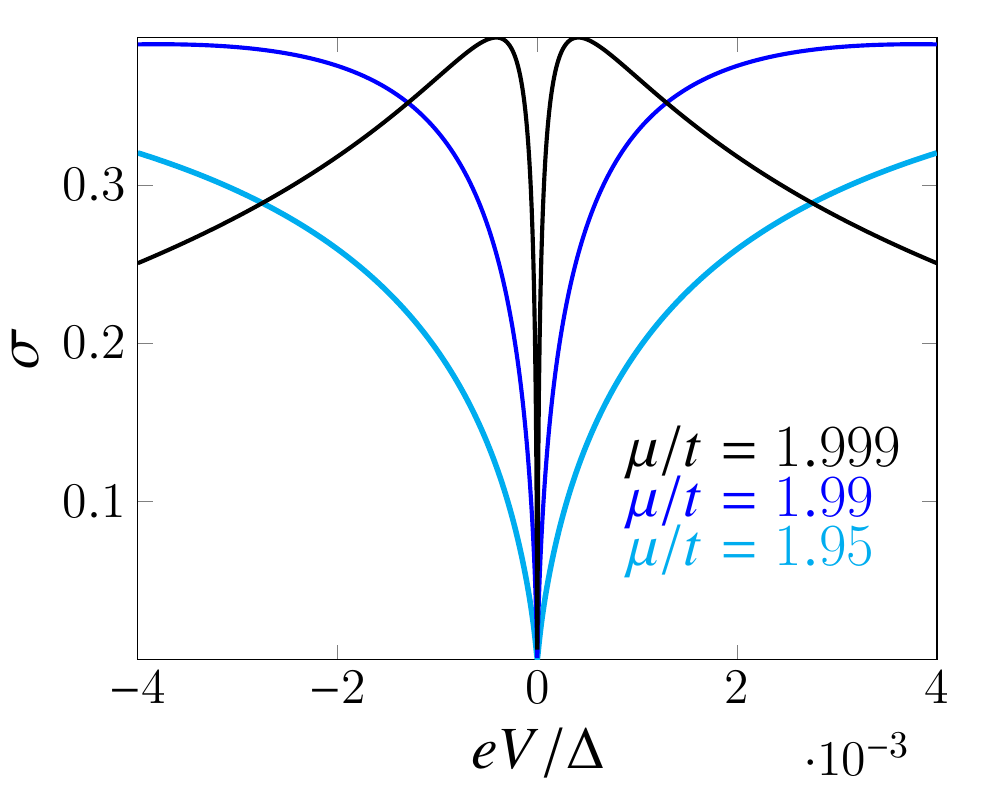}
    \caption{
        Standard deviation $\sigma$ of differential conductance in the normal metal/diffusive normal metal/Kitaev chain junction is plotted as a function of bias voltage $eV$.
        It is independent of $j$ in the normal metal region.
        $\mu_\mathrm{N}/t =0.5$, $\Delta/t=0.1$, $\delta_\epsilon/t=10^{-8}$,
        $j=-10$, and $W/t=1$.
        Topological regime near topological critical point: $\mu = 1.99$, $1.95$, $1.999$.
    }
    \label{fg:cond-sigma-inf}
\end{figure}

\section{Normal Metal/Diffusive Normal Metal/Kitaev Chain Junction}
\label{sect:N/DN/KC}

To study the effect on impurities in the normal metal/diffusive normal metal/Kitaev chain junction in \sect{\ref{sect:junction}},
we calculate the half-width at half maximum of the zero-energy peak of the local density of states for the length $L$ of the diffusive normal metal [\fig{\ref{fg:ldos-amp}(a)}]
and for the intensity $W$ of impurity scattering [\fig{\ref{fg:ldos-amp}(b)}].
These results show that the half-width at half maximum of the zero-energy peak is proportional to $\exp(c_1 L + c_2 W)$.

To clarify the perfect resonance nature,
we calculate the standard deviation $\sigma$ of differential conductance $G_\mathrm{NS}$:
\begin{align}
    \sigma = \sqrt{\frac{1}{\rmax}\sum_{l=1}^{\rmax} \left[G_\mathrm{NS}^{(l)} - G_\mathrm{NS}^{\mathrm{(avg.)}}\right]^2},
\end{align}
where $G_\mathrm{NS}^l$ is the differential conductance calculated at the $l$th times and 
$G_\mathrm{NS}^\mathrm{(avg.)}$ is the average of the differential conductance with the sample impurities.
The resulting standard deviation
for various $\mu$ is shown in \fig{\ref{fg:cond-sigma-inf}}. 
The standard deviation is always zero at $eV = 0$ in the topological regime. 
This result means the absence of shot noise and strong resonance occurs independently of $L$ \cite{Pablo3}.
From the above, it is found that the zero-bias conductance is robust against impurity scattering within the numerical accuracy.

\bibliographystyle{apsrev4-1}
\bibliography{TopologicalSC}

\end{document}